\documentclass[notitlepage,nofootinbib,showpacs,unsortedaddress,aps,prd,12pt,sort&compress]{revtex4-1}
\usepackage[utf8x]{inputenc}
\usepackage{amsmath, graphicx, amsfonts, slashed}
\usepackage{hyperref}

\newcommand{\be}{\begin{equation}}
\newcommand{\ee}{\end{equation}}
\renewcommand{\eqref}[1]{Eq.~(\ref{#1})}

\begin{document}

\title{Analytic and numerical study of the free energy in gauge theory}

\author{Axel Maas}\thanks{Supported by the DFG under grant number MA 3935/5-1}
 \email{axelmaas@web.de}
 \affiliation{Institute for Theoretical Physics,
              Friedrich-Schiller-University Jena, Max-Wien-Platz 1, D-07743 Jena, Germany}

\author{Daniel Zwanziger}
\email{dz2@nyu.edu}
\affiliation{Physics Department, New York University, 4 Washington Place, New York, NY 10003, USA}

\begin{abstract}
\noindent{\bf Abstract:} 
We derive some exact bounds on the free energy $W(J)$ in an SU(N) gauge theory, where $J_\mu^b$ is a source for the gluon field $A_\mu^b$ in the minimal Landau gauge, and $W(J)$ is the generating functional of connected correlators, $\exp W(J) = \langle \exp(J, A) \rangle$.  We also provide asymptotic expressions for the free energy $W(J)$  at large $J$ and for the quantum effective action $\Gamma(A)$ at large $A$.  We specialize to a source $J(x) = h \cos(k\cdot x)$ of definite momentum $k$ and source strength $h$, and study the gluon propagator $D(k, h)$ in the presence of this source.  Among other relations, we prove $\int_0^\infty dh \ D(k, h) \leq \sqrt 2 k$, which implies $\lim_{k \to 0} D(k, h) = 0$, for all positive $h > 0$.  Thus the system does not respond to a static color probe, no matter how strong.  Recent lattice data in minimal Landau gauge in $d =$ 3 and 4 dimensions at $h = 0$ indicate that the gluon propagator in the minimum Landau gauge is finite, $\lim_{k \to 0} D(k, 0) > 0$.  Thus these lattice data imply a jump in the value of $D(k, h)$ at $h = 0$ and $k = 0$, and the value of $D(k, h)$ at this point depends on the order of limits.  We also present numerical evaluations of the free energy $W(k, h)$ and the gluon propagator $D(k, h)$ for the case of SU(2) Yang-Mills theory in various dimensions  which support all of these findings.
\end{abstract}
\maketitle

\newcommand{\beq}{\be}
\newcommand{\eeq}{\ee}
\newcommand{\p}{\partial}
\newcommand{\beqa}{\begin{eqnarray}}
\newcommand{\eeqa}{\end{eqnarray}}

\section{Introduction}

The free energy, and its Legendre transform, the quantum effective action, play a central role in quantum field theories. As the generating functionals of correlation functions, their knowledge, in principle, grants access to all there is to know about a theory. It is often assured that these functionals are analytic in their external sources, at least away from phase transitions, and thus their derivatives yield, in a well-defined manner, the correlation functions.

However, recently this assumption yielded colliding results: Assuming this analyticity, the minimal Landau gauge gluon propagator has necessarily to vanish at zero (Euclidean) momentum \cite{Zwanziger:1991,Zwanziger:2012}. At the same time, lattice calculations, which do not need external sources, found this propagator to be finite, at least in three and four dimensions\footnote{The situation in two dimensions is \cite{Maas:2007uv,Cucchieri:2011ig} because of kinematic reasons, different, and there the propagator always vanishes \cite{Zwanziger:2012,Cucchieri:2012cb,Huber:2012zj}.} \cite{Cucchieri:2007rg,Cucchieri:2007md,Cucchieri:2010xr,Bogolubsky:2007ud,Bogolubsky:2009dc,Bornyakov:2009ug} for a review. At the same time, continuum calculations using functional methods, where the functional equations were derived under this assumption, found both solutions \cite{Fischer:2008uz,Boucaud:2008ji,Binosi:2009qm,Dudal:2008sp,Vandersickel:2012tz}. See \cite{Maas:2011se} for a review of the situation.

The logical starting point to resolve this discrepancy is therefore to check the analyticity of the free energy. This is the aim in this work.

To this end, we shall be concerned with the Euclidean correlators of gluons in QCD with an arbitrary gauge group, here chosen to be $SU(N)$, for the local gauge symmetry that are fixed to the minimal Landau gauge. These are the fundamental quantities in quantum field theory.

The minimal Landau gauge \cite{Maas:2011se} is obtained by minimizing the Hilbert square norm
\beq
|| A ||^2 = \int | A_\mu^b(x) |^2 d^4x,
\eeq  
to some local minimum (in general not an absolute minimum) with respect to local gauge transformations $g(x)$.  These act according to ${^g}A_\mu = g^{-1} A_\mu g + g^{-1} \p_\mu g$.  At a local minimum, the functional $F_A(g) \equiv ||{^g}A||^2$ is stationary and its second variation is positive.  It is well known that these two properties imply respectively that the Landau gauge (transversality) condition is satisfied, $\p \cdot A = 0$, and that the Faddeev-Popov operator is positive {\it i. e.} $(\omega, M(A) \omega) \geq 0$ for all $\omega$.  Here the Faddeev-Popov operator acts according to
\beq
\label{Macts}
M^{ac}(A) \omega^c = -  D_\mu^{ac}(A) \p_\mu \omega^c,
\eeq
where the gauge covariant derivative is defined by $D_\mu^{ac}(A) \omega^c = \p_\mu \omega^a + f^{abc} A_\mu^b \omega^c$, and the coupling constant has been absorbed into $A$.  Configurations $A$ that satisfy these two conditions are said to be in the (first) Gribov region \cite{Gribov:1977wm} which we designate by $\Omega$.  It is known \cite{Maas:2011se} that in general there are more than one local minimum of $F_A(g)$, and we do not specify which local minimum is achieved.  This gauge is realized numerically by minimizing a lattice analog of $F_A(g)$ by some algorithm, and the local minimum achieved is in general algorithm dependent. However, for all commonly employed algorithms this does not yield different expectation values, as they all are equivalent to an averaging over the first Gribov region \cite{Maas:2011se,Maas:2013vd}.

The analytic bounds which we shall obtain in section \ref{sanalytic} follow from the restriction of the gauge-fixed configurations to the Gribov region $\Omega$, and are the same whether the gluons are coupled to quarks as in full QCD, or not, as in pure gluodynamics.  In fact the same bounds hold for other gauge bosons with $SU(N)$ gauge symmetry, for example, in the Higgs sector, provided only that the gauge-fixing is done to the minimal Landau gauge. The impact of these results on the aforementioned discrepancy on the gluon propagator from different methods is then discussed in section \ref{spw}. Finally, the numerical results we shall present in section \ref{snumeric} will be for pure gluodynamics in $SU(2)$ gauge theory. 

Some preliminary results on this topic were presented in \cite{Zwanziger:2012pw,Maas:2013mz}.

\section{Bounds and other properties of free energy and quantum effective action}\label{sanalytic}

\subsection{Optimum bound on free energy}\label{soptbound}

In the minimal Landau gauge, the free energy $W(J)$ is defined by
\beq
 \exp W(J) \equiv \int_\Omega dA \ \rho(A) \ \exp(J, A).\label{freeenergy}
 \eeq
Here any matter degrees of freedom (if present) are integrated out.  The Euclidean probability $\rho(A)$ includes the Yang-Mills action, the gauge-fixing factor $\delta(\p \cdot A)$, the Faddeev-Popov determinant, and possibly the matter determinant.  We shall use only the properties $\rho(A) \geq 0$ and $\int_\Omega dA \ \rho(A) = 1$, and that the region of integration is bounded in every direction \cite{Zwanziger:1982}.  The source term $J_\mu^a(x)$ is real and is taken to be transverse $\p \cdot J = 0$ without loss of generality because $A$ is identically transverse.  The free energy per unit Euclidean volume, $w(J) = W(J)/V,$ is the generating functional of connected correlators,
\beq
\langle A(x) A(y) ... \rangle_{\rm conn} = {\p \over \p J_x}  {\p \over \p J_y} ... w(J).\label{genfunc}
\eeq

The general bound is immediate.  From the inequality $(J, A) \leq \max_{A \in \Omega} (J, A) = (J, A^*)$, where $A^*(J)$ is that configuration in $\Omega$ that maximizes $(J, A)$ for fixed $J$, we obtain
\beqa
\exp W(J) & \leq &  \int_\Omega dA \ \rho(A) \  \exp(J, A^*) \nonumber \\
& = & \exp(J, A^*),
\eeqa
which gives the bound
\beq
W(J) \leq (J, A^*(J)).
\eeq
Because $\Omega$ is bounded in every direction \cite{Zwanziger:1982}, this bound is finite.\footnote{``Bounded in every direction" means that for any configuration $A_{x, \mu}^b \neq 0$ in $\Omega$, there exists a positive number $\sigma$ such that the, not necessarily gauge-equivalent, configuration $\sigma A_{x, \mu}^b$  lies outside $\Omega$.}  Moreover the maximum configuration $A^*$ must lie on the boundary $\p\Omega$.  Indeed let $A \in \Omega$ be an interior point of $\Omega$, then for some parameter $\lambda$ sufficiently close to 1, the configuration $A' = \lambda A$ is also in $\Omega$ and $(J, A') = \lambda (J, A) > (J, A)$, so $A \neq A^*$.  This gives the more precise bound on the free energy,
\beq
\label{genbound}
W(J) \leq \max_{A \in \p \Omega}(J, A) = (J, A^*(J)),
\eeq
where $A^*(J)$ maximizes $(J, A)$ for all $A^* \in \p \Omega$.  This is the optimum bound, if all that is known is that $\rho(A)$ vanishes outside $\Omega$.   The right hand side is linear in $J$,
\beq
\max_{A \in \p \Omega}( h J, A) = h \max_{A \in \p \Omega}(J, A)
\eeq
for $h > 0$.  This is in stark contrast to a free theory which is quadratic in $J$,
\beq
W_{\rm free}(J) = (1/2) (J, K^{-1}J),
\eeq
where $K = - \p^2 +  m^2$, which strongly violates the linear bound (\ref{genbound}) at large $J$.  The linear bound on the free energy is a direct consequence of the existence of the Gribov horizon in gauge theories.\footnote{Note that the argument so far is actually not specific to gauge theories; it is sufficient that the relevant field fluctuations are bounded. It thus also applies, e.\ g., to non-linear $\sigma$-models with a positive metric target space.}

\subsection{Asymptotic free energy}

If we add the information that $\rho(A)$ is strictly positive, $\rho(A) > 0$ for all $A \in \Omega$, then the optimum bound is saturated,
\beq
\lim_{h \to + \infty}W(h J) = h \ {\rm max}_{A \in \p\Omega}(J, A) +o(h),
\eeq
where the remainder is subdominant, $\lim_{h \to \infty} o(h)/h = 0$.  Indeed this follows from
\beq
 \lim_{h \to + \infty} \exp W(hJ) =  \lim_{h \to + \infty} \int_\Omega dA \ \rho(A) \exp[h(J, A)],
 \eeq
upon taking the limit $h \to + \infty$ at fixed $J$.  Thus the asymptotic free energy, defined by
\beq
\label{maxJA}
W_{\rm as}(J) = (J, A^*(J)).
\eeq
is finite and linear $W_{\rm as}(h J) = h W_{\rm as}(J)$ for $h > 0$, and we may express the inequality (\ref{genbound}) at finite $J$ as
\beq
W(J) \leq W_{\rm as}(J).
\eeq

   Here we have assumed that $\rho(A)$ is strictly positive $\rho(A) > 0$ for all $A$ in $\Omega$.  Suppose now that this is not the case, and that the numerical gauge fixing is such that $\rho(A)$ is strictly positive only on a proper subset $\Lambda \subset \Omega$.  Each set has its own free energy $W_\Lambda(J)$, its asymptotic free energy,
\beq
\label{limWLambda}
\lim_{h \to +\infty} W_\Lambda(h J) = h W_{\Lambda,{\rm as}}(J),
\eeq
which is linear in $J$, and given by
\beq
\label{WLambdaas}
W_{\Lambda, {\rm as}}(J) = \max_{A \in \Lambda} (J, A) \leq \max_{A \in \Omega} (J, A) = W_{\Omega, {\rm as}}(J),
\eeq
Thus the inequalities,
\beq
\label{Lambdasubset}
W_\Lambda(J) \leq W_{\Lambda, {\rm as}}(J) \leq W_{\Omega, {\rm as}}(J) \ {\rm for} \ \Lambda \subset \Omega,
\eeq
hold, for each $J$.

\begin{figure}
\includegraphics[width=\linewidth]{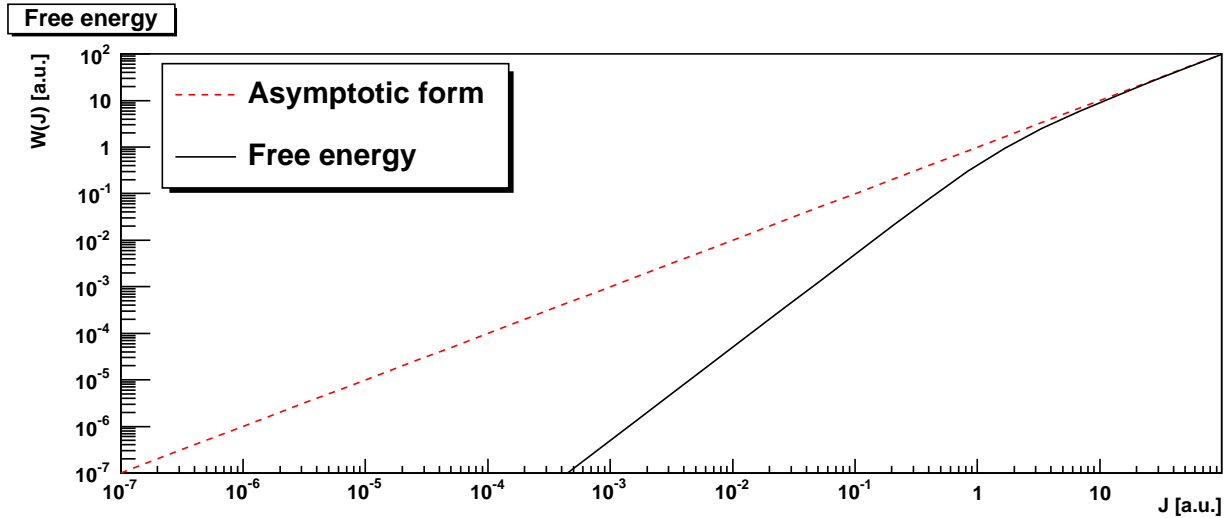}\\
\caption{\label{fig:was-illu}An illustration of how the true free energy could approach the asymptotic limit (\ref{limWLambda}).}
\end{figure}

This situation is illustrated in figure \ref{fig:was-illu}, where it is shown that the free energy, always saturated by the bound, will ultimately approach the asymptotic form from below.

\subsection{Bound on the approximate free energy $W_{\rm num}(J)$ obtained from a finite Monte Carlo process}\label{sanamc}

We shall report below on a numerical determination of the free energy $\exp[W(J)] = \langle \exp(J, A) \rangle$, but let us first discuss what kind of limitations we expect in this case.

In the numerical determination, a finite number of configurations $A_n$, $n = 1, ... N$ is generated by a Monte Carlo process, and the approximate free energy, $W_{\rm num}(J)$, obtained from the Monte Carlo process, is given by the average over the sample set $\Sigma = \{ A_n \}$,
\beq
\exp[W_{\rm num}(J)] = N^{-1} \sum_n \exp(J, A_n).
\eeq
Each gauge-fixed configuration $A_n$ is obtained as described in the introduction, so each configuration $A_n$ lies inside the Gribov region $\Omega$, and the sample set ${\Sigma \equiv \{ A_n} \}$, is a subset of the Gribov region $\Sigma \subset \Omega$.  Consequently eqs.~(\ref{limWLambda}) to  (\ref{Lambdasubset}) hold, with $\Lambda \to \Sigma$.  Thus the asymptotic limit of the approximate free energy,
\beq
\label{limWnum}
\lim_{h \to +\infty} W_{\rm num} (h J) = h W_{\rm num,as}(J),
\eeq
is given by
\beq
\label{Wnumas}
W_{\rm num,as}(J) = \max_n (J, A_n).
\eeq
It is linear in $J$.  Moreover the inequalities,
\beq
\label{numsubset}
W_{\rm num}(J) \leq W_{\rm num,as}(J) \leq W_{\Omega, {\rm as}}(J),
\eeq
hold, for each $J$.  

This implies the following observations:
\begin{enumerate}
\item The asymptotic form $W_{\rm num,as}(J)$ of the approximate free energy obtained from the Monte Carlo process is linear in $J$, {\em no matter how inaccurate the numerical determination may be.}  In the extreme case of only one sample configuration $A_1$, then $W_{\rm num}(J) = W_{\rm num,as}(J) = (J, A_1)$, which is indeed linear in~$J$.
\item According to the inequality, $W_{\rm num,as}(J) \leq W_{\Omega, {\rm as}}(J)$,  {\em any inadequacy of the sample set~$\{ A_n\}$ can only result in an undersaturation of the optimal bound,} $W_{\Omega, {\rm as}}(J)$.  This accords with the intuition that if there is no configuration $A_n$ in the numerical sample that is sufficiently close to $A^*(J)$, which lies on the Gribov horizon $\p \Omega$, this could result in significant undersaturation of the optimal bound. 
\item The asymptotic form of the approximate free energy $W_{\rm num}$(J) may be calculated directly from \eqref{Wnumas}.
\end{enumerate}

We now discuss the (possible) undersaturation further, according to which the sample configurations are not close to the Gribov horizon.  This might seem surprising since it is often considered that the probability distribution is concentrated close to the Gribov horizon.   To address this question, we consider a toy model.  Suppose that the configurations are parametrized by coordinates $a_i$ for $i = 1, ... N$, where $N$ is a large number.  In a lattice theory $N$ would be of order of the lattice volume $V$.  Suppose that the Gribov region is the sphere $\sum_i a_i^2 \leq N$.  We ignore other effects, and take the measure to be 
\beq
\int \prod_{i = 1}^N da_i \ \theta\Big(N - \sum_{i = 1}^N a_i^2\Big),
\eeq
where $\theta$ is the step function. This model shares with the exact  theory the property that (1) it is convex \cite{Zwanziger:1982}, (2) it is contained within an ellipsoid, Appendix B of \cite{Zwanziger:1991} (so after the coordinates are rescaled it is contained within a sphere), and (3) the horizon function is a bulk quantity, proportional to the volume $V$.  We introduce the radial coordinate
\beq
r = \Big( \sum_{i = 1}^N a_i^2\Big)^{1/2}, 
\eeq
whose probability distribution is given by
\beq
\int_0^{N^{1/2}} dr \ r^{N-1}.
\eeq
Here and below we ignore an over-all normalization constant.  For large $N$ the probability gets concentrated at the upper limit, $r = N^{1/2}$, and the measure approaches
\beq
\int_0^\infty dr \ \delta(r - N^{1/2}).
\eeq
which is entirely concentrated on the Gribov horizon.  (This shows that at large $N$ $\delta(N - r^2)$ and $\theta(N - r^2)$ are statistically equivalent.)

Consider now the probability distribution of a single variable $a_1$, and integrate out all other variables $a_i$ for $i = 2, ... N$.  This is the quantity of interest when we consider a source term $(J, A) = N j_1 a_1$.  We introduce a radial coordinate for the remaining variables,
\beq
\rho = \left( \sum_{i = 2}^N a_i^2\right)^{1/2}. 
\eeq  
The measure is now
\beq
\int_{- N^{1/2}}^{N^{1/2}} da_1 \int_0^{(N - a_1^2)^{1/2}} d\rho \ \rho^{N-2} = (N-1)^{-1} \int_{- N^{1/2}}^{N^{1/2}} \ da_1 \ (N - a_1^2)^{(N-1)/2},
\eeq
which, for large $N$, is concentrated near the origin, $a_1 = 0$, and not at its maximum magnitude, $a_1 = \pm N^{1/2}$.  To quantify this, we use
\beqa
(N - a_1^2)^{(N-1)/2} & = & \exp[(1/2)(N-1) \ln(N - a_1^2)]
\nonumber  \\
& = & \exp \{(1/2)(N-1) [\ln N +  \ln(1 - a_1^2/N)] \}
\nonumber  \\
& \approx & \exp[(1/2)(N-1) \ln N  -  (1/2)a_1^2)],
\eeqa
which is valid at large $N$.  Thus the measure of a single variable, if the others are integrated out, is
\beq
\int da_1 \exp( - a_1^2/2 ).
\eeq
This is a Gaussian measure of unit width, whereas the maximum magnitude $a_1$ may attain inside the Gribov horizon is, $a_1 = \pm N^{1/2}$, where $N$ is a large number.  It is highly unlikely that the Monte Carlo process will sample $a_1$ close to the Gribov horizon, and we should therefore expect under-saturation.  As we have just seen, this happens by the same mechanism as equipartition of energy in statistical physics: it is highly unlikely that a single given molecule of the air inside a room would carry all the kinetic energy, leaving zero kinetic energy for every other air molecule.  Although a ``typical" configuration will indeed be close to the Gribov horizon, as measured by the variable $r^2$ which is the ``horizon function" for this toy model, nevertheless it is highly improbable that a single given variable will be close to its maximum allowed value, the other variables then being constrained close to 0.

\subsection{Formula for $A^*(J)$}

The boundary $\p \Omega$ is described by the equation $\lambda_0(A) = 0$, where $\lambda_0(A)$ is the lowest non-trivial eigenvalue\footnote{The trivial eigenvalue belongs to constant eigenfunctions which satisfy $\p_\mu \omega = 0$.} of the Faddeev-Popov operator $M(A) = - D_\mu(A)\p_\mu$.  Here we stipulate that the Euclidean volume $V = L^d$ is finite (though large), so eigenvalues are discrete, and eigenfunctions normalizable,
\beq
\p \Omega = \{A: \lambda_0(A) = 0\}.
\eeq
According to the Lagrange multiplier method, we may find the point $A^*$ which maximizes $(J, A)$ for $A \in \p\Omega$ by finding the stationary point $A^*$ of
\beq
\label{Lagrangefunct}
I(A) = (J, A) -  \alpha \lambda_0(A).
\eeq
It is the solution of 
\beq
\label{lagrangeeq}
{\delta I(A) \over \delta A_{x, \mu}^b} = J_{x,\mu}^b - \alpha {\delta  \lambda_0(A) \over \delta A_{x, \mu}^b} = 0,
\eeq
where the Lagrange multiplier $\alpha$ is determined by 
\beq
 \lambda_0(A) = 0.
\eeq
As we have seen, the asymptotic free energy is then given by $W_{\rm as}(J) = (J, A^*)$.

Geometrically (\ref{lagrangeeq}) states that $J$ lies along the normal
\beq
N_{x, \mu}^b = {\delta  \lambda_0(A) \over \delta A_{x, \mu}^b},
\eeq
to the Gribov horizon which is the surface defined by $\lambda_0(A) = 0$.  There is a simple explicit formula for this normal.  Let $\psi_0 = \psi_0(A)$ be the normalized eigenfunction belonging to $\lambda_0(A)$, so $M(A) \psi_0 = \lambda_0(A) \psi_0$, and
\beq
\label{eigenmean}
\lambda_0(A) = (\psi_0(A), M(A) \psi_0(A)).
\eeq
There is a theorem due to Feynman\footnote{To prove Feynman's theorem, we consider a small variation $\delta A$ of $A$ in \eqref{eigenmean},
\beq
\nonumber
\delta\lambda_0(A) = (\psi_0(A), \delta M(A) \psi_0(A)) + R,
\eeq
where
\beq
\nonumber
R \equiv  (\delta \psi_0(A), M(A) \psi_0(A)) +  (\psi_0(A), M(A) \delta\psi_0(A)).
\eeq
It is sufficient to prove that $R$ vanishes.  We have
\beqa
R & = & \lambda_0 [ (\delta \psi_0(A), \psi_0(A)) +  (\psi_0(A), \delta\psi_0(A)) ] 
\nonumber \\
\nonumber
& = &  \lambda_0 \delta (\psi_0, \psi_0) = 0 ,
\eeqa
which vanishes because $\psi_0(A)$ is normalized $(\psi_0(A), \psi_0(A)) = 1$.  QED} which asserts that in the last formula the variation of $\lambda_0(A)$ results only from the variation of the operator $M(A)$ of which it is the eigenvalue, but not from the variation of the eigenfunction $\psi_0(A)$,
\beq
{\delta  \lambda_0(A) \over \delta A_\mu^b(x)} = \left(\psi_0(A), {\delta  M(A) \over \delta A_\mu^b(x)} \psi_0(A)\right).
\eeq
With $M(A)$ given in \eqref{Macts} we obtain
\beq
{\delta  \lambda_0(A) \over \delta A_\mu^b(x) } = - f^{abc} \psi_0^{a*}(x) \p_\mu \psi_0^c(x).
\eeq
With this result, \eqref{lagrangeeq} may be written, after a simple rescaling,
\beq
\label{Jnormal}
J_\mu^a(x) =  [f^{abc} \psi_0^{b*}(x) \p_\mu \psi_0^c(x)]^{\rm tr}.
\eeq
Note that $A$ is identically transverse $A = A^{\rm tr}$, so only the transverse part of $J$ is operative in $(J, A)$, and one may impose transversality on $J$ as is done here.  One may verify that the transverse source $J$ given here is purely real.

In general, given $J$, it is difficult to find $A^*(J)$.  However we may solve the inverse problem.  We find a configuration $A^*$ that lies on the Gribov horizon, by solving the eigenvalue problem $M(A^*) \psi_0 = \lambda_0(A^*) \psi_0 = 0$.  Then $J$, given by \eqref{Jnormal}, is normal to the Gribov horizon at $A^*$.  This provides $W_{\rm as}(J) =  (J, A^*)$.  This method will be used in Appendix A for the source $J$ we shall study numerically.

\subsection{Cusp in Gribov horizon}\label{cusp}

In the last section it was assumed that the point $A^*$ on the Gribov horizon $\p \Omega$ is a regular point in the sense that the normal $N(A^*)$ at $A^*$ is unique.  However it may happen that the point $A^*$ is a cusp \cite{Greensite:2004ur}, see below, Fig.\ \ref{fig:example}.  In this case the normal is not unique, and there is a continuum of planes through $A^*$, but which otherwise lie outside $\Omega$.  Let $J$ be normal to such a plane.  Then $A^*$ maximizes $(J, A^*)$ for $A \in \p \Omega$, and as before, we have $W_{\rm as}(J) =  (J, A^*)$.  The case that we shall investigate numerically is precisely of this type.

\subsection{Asymptotic quantum effective action}\label{sqea}

In the determination of connected correlation functions the quantum effective action $\Gamma(A)$ plays a central role. It is obtained by Legendre transformation from $W(J)$
\beq
\Gamma(A) = (J, A) - W(J)\label{legendre}
\eeq
where		 
\beq
A_x(J) = { \p W(J) \over \p J_x }.
\eeq
Here the discrete index $x$ represents position and color and Lorentz indices.  $A_x(J)$ is the expectation value in the presence of the source $J$,
\beqa
\label{average}
A_x(J) & = & \langle A_x \rangle_J
\nonumber   \\
& = & { \int_\Omega dA \ \rho(A) \exp(J, A) \ A_x \over \int_\Omega dA \ \rho(A) \exp(J, A)},
\eeqa
which is an average over $\Omega$ with a positive weight $\rho(A) \exp(J, A)$.\footnote{Here for purposes of discussion we assume the probability $\rho(A)$ is non-zero out to the boundary of the Gribov region~$\Omega$, where the boundary is given by $\lambda_0(A) = 0$.  If $\rho(A)$ is non-zero over a more restrictive region, such as the fundamental modular region $\Lambda(A)$, then, more generically, the boundary is denoted by $H(A) = 0$, and $\lambda_0(A)$ gets replaced by $H(A)$ in the following discussion.}  Clearly the average over the convex region $\Omega$ lies inside the convex region that is averaged over
\beq
A_x(J) \in \Omega,
\eeq
so the domain of definition of $\Gamma(A)$ is the Gribov region~$\Omega$.  Moreover for finite $J$ this average over $\Omega$ yields an interior point of $\Omega$, and a boundary point is approached only for $J \to \infty$.  As before, the integrals in (\ref{average}), for $J = h \widehat J$ are dominated at large $h$ with $\widehat J$ fixed by the point $A^*(\widehat J)$ that maximizes $(\widehat J, A)$ for points $A \in \p \Omega$,
\beq
\label{AtoAstar}
\lim_{h \to \infty}A_x(h \widehat J) = A_x^*(\widehat J) \in \p \Omega.
\eeq

If we attempt to calculate the asymptotic form $\Gamma_{\rm as}(A)$ of the quantum effective action directly from the  asymptotic form $W_{\rm as}(J)$, we get
\beq
\Gamma_{\rm as}(J) = \Big(J, {\p W_{\rm as} \over \p J } \Big) - W_{\rm as}(J) = 0,
\eeq	
because $W_{\rm as}(J)$ is linear in $J$,  $W_{\rm as}(h J) = h W_{\rm as}(J)$,
so by Euler's equation,
\beq
 \sum_x J_x {\p W_{\rm as} \over \p J_x } = W_{\rm as}(J).
\eeq

However the situation is not hopeless.  The inverse Legendre transformation reads
\beq
J_x(A) = {\p \Gamma(A) \over \p A_x},
\eeq
and, by \eqref{AtoAstar}, if $A$ approaches a boundary point $A \to A^* \in \p \Omega$, then $J(A)$ diverges like $J(A) = h \widehat J(A^*)$ with $h \to \infty$ and $\widehat J(A^*)$ fixed.  Thus as $A$ approaches the boundary point $A^*$, the quantum effective action approaches an asymptotic form,  $\Gamma(A) \approx \Gamma_{\rm as}(A)$ that satisfies
\beq
\label{taudiverges}
h \widehat J(A^*) \approx {\p \Gamma_{\rm as}(A) \over \p A_x },
\eeq
where $h$ diverges as $A$ approaches $A^* \in \p\Omega$.  With $J = h \widehat J$, we have also seen, \eqref{lagrangeeq}, that, for $A$ approaching a boundary point, $A \approx A^*$,
\beq
h \widehat J(A^*) \approx \alpha {\p \lambda_0(A) \over \p A_x }.
\eeq
This gives
\beq
\label{parallel}
{\p \Gamma_{\rm as}(A) \over \p A_x } \approx \alpha(A) {\p \lambda_0(A) \over \p A_x },
\eeq
and so, for $A$ approaching a boundary point $A^* \in \p\Omega$, the normal to a surface of constant $\Gamma_{\rm as}(A)$ is also everywhere normal to a surface of constant $\lambda_0(A)$.  Thus the surfaces of constant $\Gamma_{\rm as}(A)$ coincide with the surfaces of constant $\lambda_0(A)$, and so, as $A$ approaches the boundary, the asymptotic quantum effective action $\Gamma_{\rm as}(A)$ depends on $A$ only through $\lambda_0(A)$,
\beq
\Gamma_{\rm as}(A) = f[\lambda_0(A)],
\eeq
where $f(\lambda_0)$ is an unknown function, but depending only on the single variable~$\lambda_0(A)$.  This formula is valid for $A$ approaching a boundary point $A^*$.  We have
\beq
{\p \Gamma_{\rm as}(A) \over \p A_x } = f'(\lambda_0(A)) {\p \lambda_0(A) \over \p A_x }.
\eeq
Moreover $f'(\lambda_0)$ diverges as $A$ approaches a boundary point $A^*$,
\beq
f'(\lambda_0(A^*)) = f'(0)  = \infty,
\eeq
because in (\ref{taudiverges}) $h$ diverges for $A \to A^*$.

  We now add a speculation to the above reasoning.  Observe that the Faddeev-Popov determinant is the product of eigenvalues, $\det[M(A)] = \exp[- S_{\rm eff}(A)] = \prod_n \lambda_n$, so $S_{\rm eff}(A) = - \sum_n \ln \lambda_n$, and in the semi-classical limit $\Gamma(A) \sim S_{\rm eff}(A)$.  This suggests that $\Gamma_{\rm as}(A) = f(\lambda_0)$, is given by
\beq
\Gamma_{\rm as}(A) = - V \hat\gamma \ln \lambda_0(A).
\eeq
Here $V = L^d$ is the Euclidean volume, which is required because $\Gamma$ is a bulk quantity, and $\hat\gamma$ is a constant of dimension $(\rm mass)^d$.

We shall shortly show that for the configuration $A_\mu^b(x) = c \cos(kx_1) \delta_{\mu 2} \delta^{b3}$, the lowest eigenvalue is given by $\lambda_0 = (2 \pi /L)^2(1 - c^2/2k^2)$.
It is more convenient to work with normalized basis $\sqrt 2 \cos(k x_1)$, and we write $c = \sqrt 2 \hat c$.  Our asymptotic formula is then given by
\beq
\gamma_{\rm as}(\hat c) \equiv \Gamma_{\rm as}(\hat c)/V = - \hat\gamma \ln(1 - \hat c^2/k^2).\label{qefactspec}
\eeq
The free energy per unit Euclidean volume corresponding to this expression is given by the Legendre transformation,
\beq
w(\hat h) = \hat h \hat c - \gamma_{\rm as}(\hat c),
\eeq 
where $\hat c = \hat c(\hat h)$ is determined by
\beq
\hat h = {\p \gamma_{\rm as}(\hat c) \over \p \hat c}.
\eeq
One finds
\beq
w(\hat h) =  \hat\gamma [1 + (\hat h k / \hat\gamma)^2]^{1/2} - \hat\gamma - \hat\gamma \ln \Big\{  { [1 + (\hat h k / \hat \gamma)^2  ]^{1/2} + 1 \over 2 } \Big\}. 
\eeq
This is the free energy of a simple model \cite{Zwanziger:2011yy}, with $\hat\gamma \to g(k)$.  This contains the leading correction at asymptotically large $\hat h$ to $w_{\rm as} = \hat h k$.

\begin{figure}
\includegraphics[width=\linewidth]{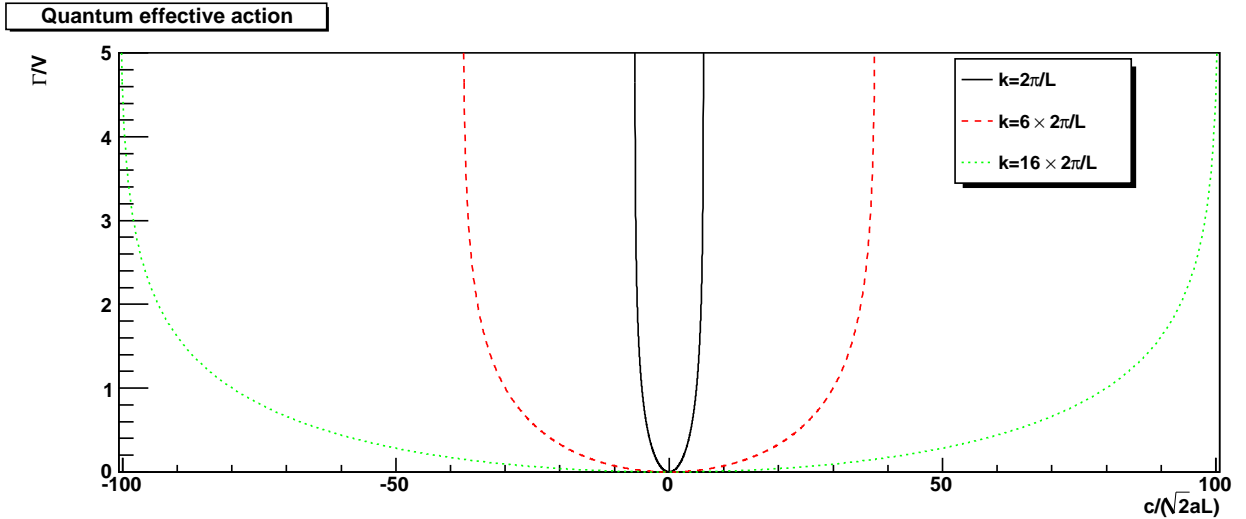}\\
\caption{\label{fig:gamma-illu}The quantum effective action (\ref{qefactspec}) for $\hat\gamma=1$. The lattice spacing is denoted by $a$.}
\end{figure}

It is a remarkable fact that the quantum effective action (\ref{qefactspec}) reflects the boundedness of the Gribov region: For each momenta $k$, there exists a finite value of the  amplitude, given by $\hat c = k$, for which the quantum effective action diverges, signaling that larger field values cannot be reached, and moreover this insurmountable barrier closes in on the origin $\hat c = 0$ as $k \to 0$. This fact is illustrated in figure \ref{fig:gamma-illu}.

\subsection{An example}\label{locatehorizon}

Before determining the free energy in a simulation, it is worthwhile to consider the concepts so far for an example. Take the configurations in $SU(2)$ gauge theory,
\beq
\label{planebc}
A_\mu(x) = [b + c \cos(kx_1)] \delta_{\mu 2} \ e_3,
\eeq
parametrized by the two real parameters $b$ and $c$.  We quantize in a periodic box of Euclidean volume $V = L^d$,  with principle axes aligned in the $x_1$- and $x_2$-directions, and $k = 2 \pi n/L$ where $n \neq 0$ is a non-zero integer.  Here $e_3$ is the unit color vector in the 3-direction, and the dependence on the Lorentz index $\mu$ and on position $x_\mu$ is chosen so these configurations are transverse $\p_\mu A_\mu = 0$.  They constitute a two-plane $P(b, c)$ through the origin in $A$-space.  The intersection of the two-plane $P(b, c)$ with the Gribov horizon $\p \Omega$ occurs where the lowest non-trivial eigenvalue of the Faddeev-Popov operator $M[A(b, c)]$ vanishes, $\lambda_0(b, c) = 0$.   

\begin{figure}
\includegraphics[width=\linewidth]{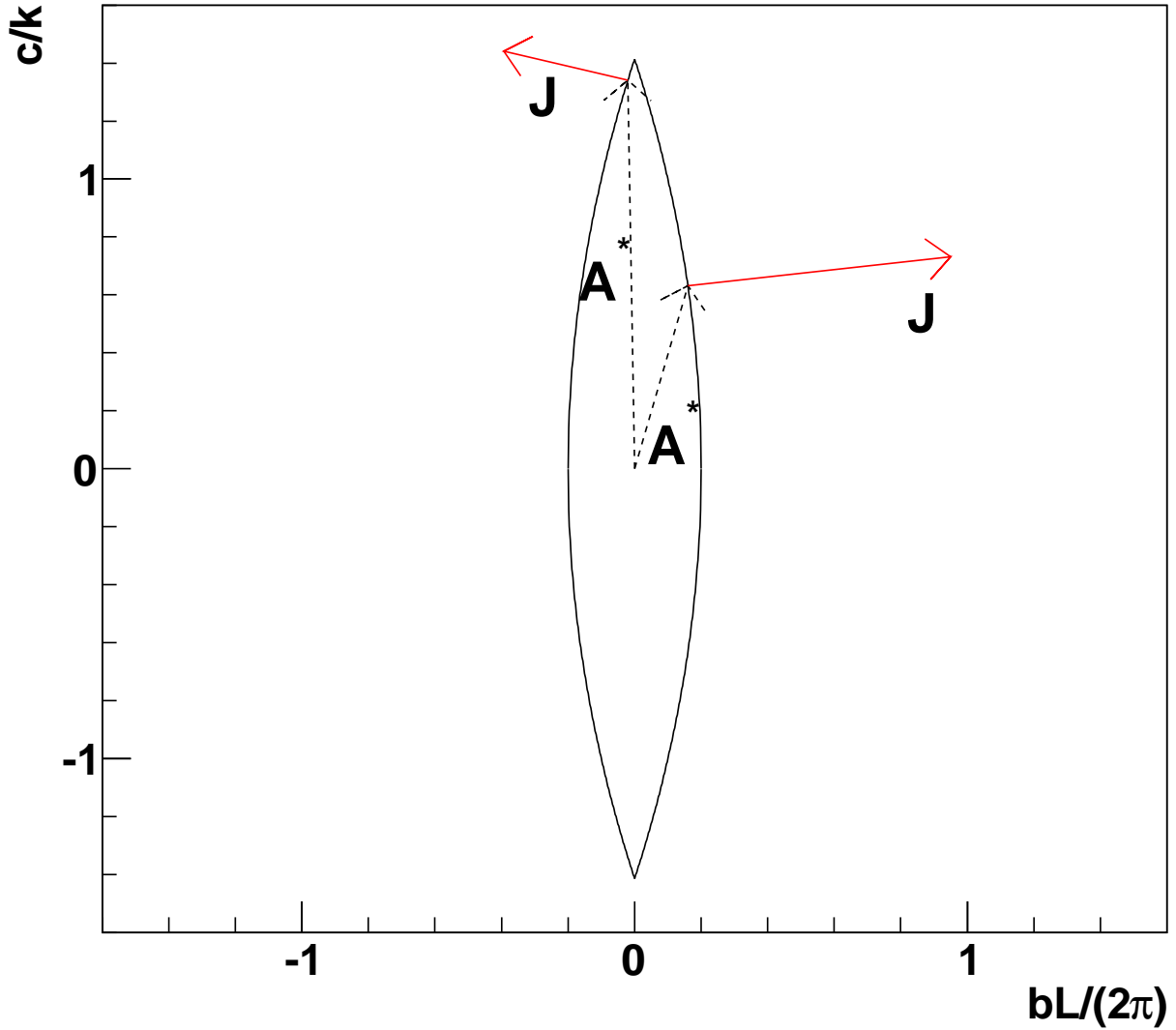}\\
\caption{\label{fig:example}The intersection of the Gribov region $\Omega$ with the 2-plane $P(b, c)$ is contained between the two parabolas.  For each $J$ there is a bound on the free energy given by $W(J) \leq (J, A^*)$, where $J$ and $A^* = A^*(J)$ are illustrated here.}
\end{figure}

 The lowest non-trivial eigenvalue, $\lambda_0(b, c)$, is calculated in Appendix \ref{locatinghorizon}, under the assumption  $k >> |p| =  2 \pi / L$, which is satisfied in the infinite-volume limit $L \to \infty$ at fixed $k$, with the result\footnote{The next order correction in $2\pi/Lk$ is given below.}
\beq
\label{lambda0}
\lambda_0(b, c) = - |p b| +  p^2 ( 1 - c^2 / 2 k^2 ) .
\eeq
The corresponding wave-function is given by
\beq
\label{psi0}
\psi_0^a(x) = [ 1 + (\rho |p| c /k^2) \cos(k x_1) ] \exp(i p x_2) \eta^a,
\eeq
where $\rho \equiv {\rm sign}(b)$, $\eta = ( e_1 - \sigma i e_2 )/\sqrt 2$ is a complex color-vector, and $\sigma = - {\rm sign}(pb)$.  The interior of the Gribov horizon is described by $\lambda_0(b, c) > 0$, and the first Gribov horizon is given by $\lambda_0(b, c) = 0$, or
\beq
b = \pm {2 \pi \over L} \Big( 1 - { c^2 \over 2 k^2 } \Big),
\eeq
This is plotted in figure \ref{fig:example}, and for $k >> 2\pi/L$, the Gribov region being the sector contained between the two rather flat parabolas.  Note the cusp at $b = 0$ and $c = \pm \sqrt 2 k$.

The configurations $A(b, c)$ for which the last equation is satisfied define points $A^*$ that lie on the Gribov horizon.  The source $J(A^*)$ that corresponds to these points, which is normal to the horizon at $A^*$ is found from (\ref{Jnormal}) and (\ref{ground state}), with the result
\beq
J_\mu(x) = |p|[ {\rm sign}(b) + (2 |p| c)/k^2) \cos(k x_1) ] \ \delta_{\mu 2} \ e_3,
\eeq
to leading order in $2\pi/L$.\footnote{There is also an imaginary component $J_1(x_1)$, which is purely longitudinal and which is eliminated when the transverse part is taken.}  The sources $J$ are also shown in figure \ref{fig:example}.

At the cusp at $A^*(b = 0, c = \sqrt 2 k)$, there are two normals in the limit $b \to 0_\pm$,  described by $J_{\mu \pm}(x) =  \pm 1  + (2 \pi /L) (2 c)/k^2) \cos(k x_1) \delta_{\mu 2} e_3$.  As discussed in sect.\ (\ref{cusp}), any linear combination of these two, that points outward from the Gribov region, also satisfies the condition $\max_{A \in \p\Omega} (J, A) = (J, A^*)$, where $A^* = A(b = 0, c = \sqrt 2 k) = \sqrt 2 k \cos(kx_1)$, and color and Lorentz indices are suppressed.  In particular the linear combination $J_\mu(x) = h \cos(kx_1) \delta_{\mu 2} e_3$ satisfies this condition.  For this source, the optimal bound, is given by
\beq
W(J) \leq (J, A^*) = \int d^dx \ h \cos(k x_1) \sqrt 2 k \cos(k x_1) = V hk /\sqrt 2.
\eeq
To this order the bound is independent of $p$, so for large volume $V = L^d$ at fixed $k$, a large number of levels $\lambda_n(A)$, all with $p_n << k$, cross through the Gribov horizon together, $\lambda_n(A^*) \approx \lambda_0(A^*)$.

The next order correction in $p^2/ k^2 = (2 \pi/L k)^2$ is given by
\beq
\lambda_0(b, c) = - |pb| + p^2(1 - c^2/2k^2) + (7/32) (p^4 c^4/k^6),
\eeq
and the Gribov horizon, $\lambda_0(b, c) = 0$, by
\beq
b = \pm |p| \  [ 1 - (c^2/2k^2) + (7/32)(p^2 c^4/k^6)].
\eeq
There is a cusp at $b = 0$, and $c = \pm \sqrt2 k \  [ 1 + (7/16) (p^2 / k^2) ]$.  For the current $J_\mu(x) = h \cos(kx_1) \delta_{\mu 2} e_3$, this gives the optimal bound
\beq
W(J) \leq (J, A^*) =  V (h k /\sqrt 2) \ [ 1 + (7/16) (p^2 / k^2) ] ,
\eeq
where $p^2 =  (2 \pi / L)^2$.

For the case $c = 0$, the Gribov horizon occurs at $ b = \pm 2\pi / L$, and the optimal bound for the source $J_\mu(x) = h \delta_{\mu 2}e_3$ is given, without approximation, by
\beq
\label{constconfig}
W(J) \leq (J, A^*) = V h (2 \pi/L).
\eeq

      We easily obtain a bound on the ``magnetization" $m(k, h) \equiv {\p w(k, h) \over  \p h} =  \langle a(k) \rangle_h$ where the last quantity is the expectation value of the Fourier component $a(k)$ of the configuration $A(x)$ in the presence of the external source $h$.  Note that $m(k, 0) = 0$, because $\langle a_k \rangle_0 = 0$ and that ${\p m(k, \ h) \over \p h} = {(1/2) D(k, h)} \geq 0$ is positive because the gluon propagator $D(k, h)$ is positive, so the magnetization $m(k, h)$ is monotonically increasing, and we have the inequalities
\beq
\label{boundmag}
0 \leq m(k, h) \leq m(k, \infty) = {\p w_{\rm as}(k, h) \over \p h } \leq k/ \sqrt 2,
\eeq
where the last inequality becomes an equality if the bound on $w(k, h)$ is saturated.  Thus in minimal Landau gauge, the magnetization produced by a static source vanishes,
\beq
\lim_{k \to 0} m(k, h) = 0,
\eeq
for all $h$, and we conclude that the static color degree of freedom cannot be excited by applying an external color-magnetic field $h$,  {\em no matter how strong.}

 \section{Gluon propagator at zero-momentum}\label{spw}\label{sgp}
 
 \subsection{Bound on gluon propagator}

To estimate further the relevance of these findings to the gluon propagator, we specialize to a plane wave source, as will be used below in section \ref{snumeric} for the numerical investigations,
 \beq
 J_\mu^a(x) = h \cos(k x_1) \delta^{a3} \delta_{\mu 2},\label{source},
 \eeq
 so
\beq
\exp[W(J)] = \Big\langle \exp[\int d^dx \ h \cos(kx_1) A_2^3(x)] \Big\rangle
\eeq
Here $h$ is the analog in a spin theory of an external magnetic field, modulated by $\cos(kx_1)$.  The wave number takes on the values $k = 2\pi n/L$, where $n$ is an integer, and $L$ is the edge of a periodic Euclidean box.  The Lorentz indices $1$ and $2$ are chosen so $J$ is transverse, $\p_\mu J_\mu = 0$.  For this source $J$ that depends on the 2 parameters $h$ and $k$, we parametrize the free energy per unit Euclidean volume $w(J) = W(J)/ V$, where $V = L^d$, by
 \beq
 w(k, h) \equiv W(J)/V.
 \eeq
 The gluon propagator is its second derivative at $h = 0$,
\beqa
 {\p^2 w(k, h) \over \p h^2} & = & (1/2) D(k, h) \ {\rm for} \ k \neq 0
 \nonumber \\ 
 & = & D(0, h) \ {\rm for} \ k = 0,
\eeqa
where we have written $D(k, h) \equiv D_{22}^{33}(k, h)$ for the gluon propagator in the presence of the source $h$.  The normalization comes from
\beqa
{\p^2 W(k, h) \over \p h^2}\Big|_{h = 0} & = & \int d^d x d^d y \cos(kx_1) \cos(ky_1) \langle A_2^3(x) A_2^3(y) \rangle_{h = 0}
\nonumber \\
& = & \int d^d x d^d y \cos(kx_1) \cos(ky_1) D_{22}^{33}(x-y),
\nonumber \\
& = &(1/2) \int d^d x d^d y \{ \cos[k (x_1 - y_1)] +  \cos[k(x_1 + y_1)] \} D_{22}^{33}(x-y),
\eeqa
where the second term does not contribute for $k \neq 0$, and we have $D(k) = D(-k) = \int d^d x \exp(i k \cdot x) D_{22}^{33}(x)$.
  
\begin{figure}
\includegraphics[width=\linewidth]{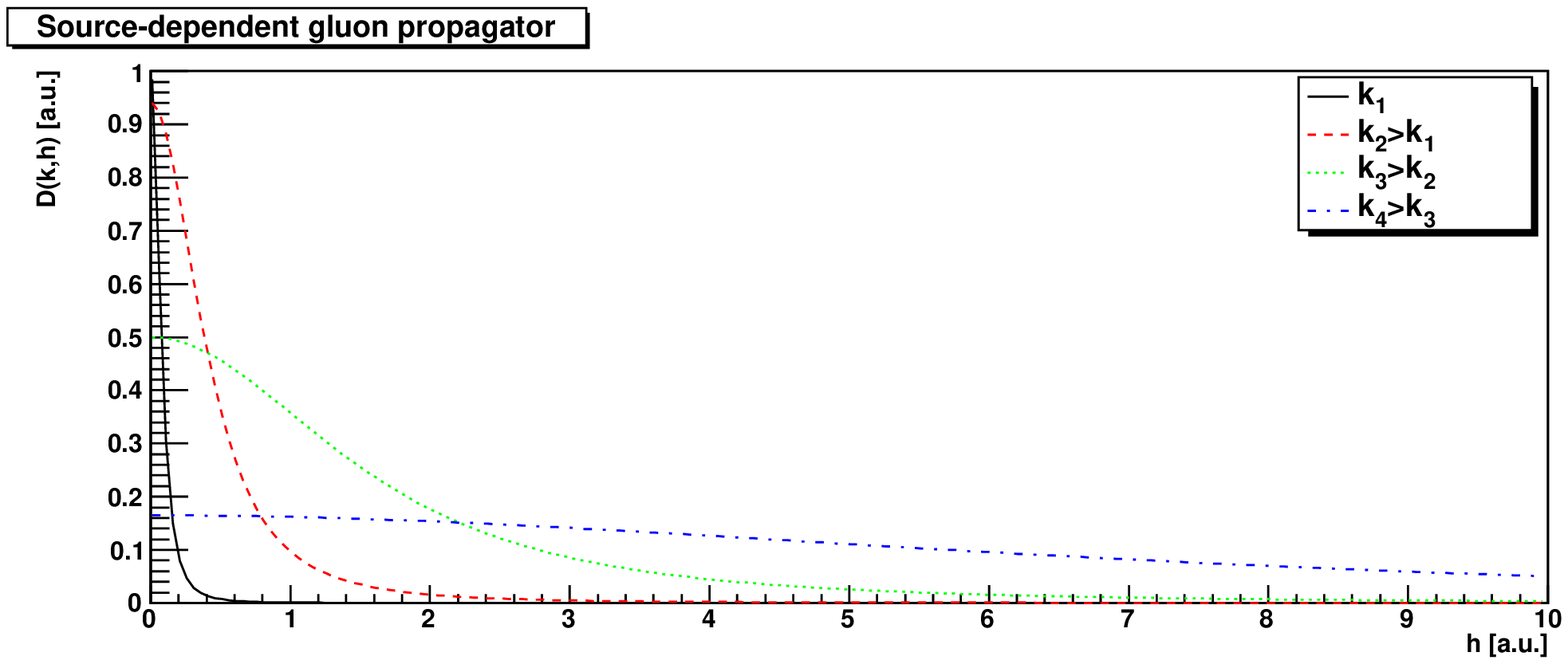}\\
\includegraphics[width=\linewidth]{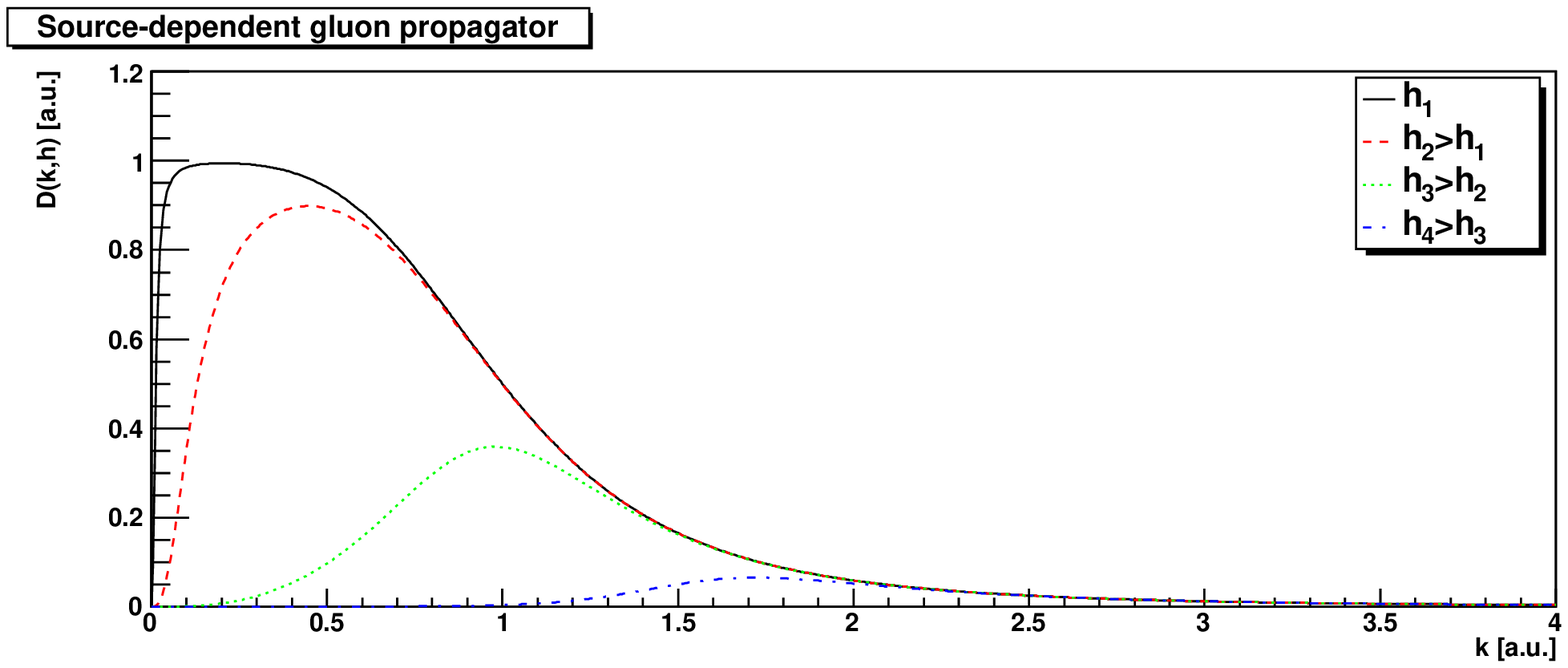}
\caption{\label{fig:d}The source-dependent infrared finite gluon propagator, (\ref{gluonprop}), with $\hat\gamma = k^2(1 + k^4)$ as a function of the source strength, for different momenta (top panel), and as a function of momentum for different source strengths (bottom panel).}
\end{figure}

\begin{figure}
\includegraphics[width=\linewidth]{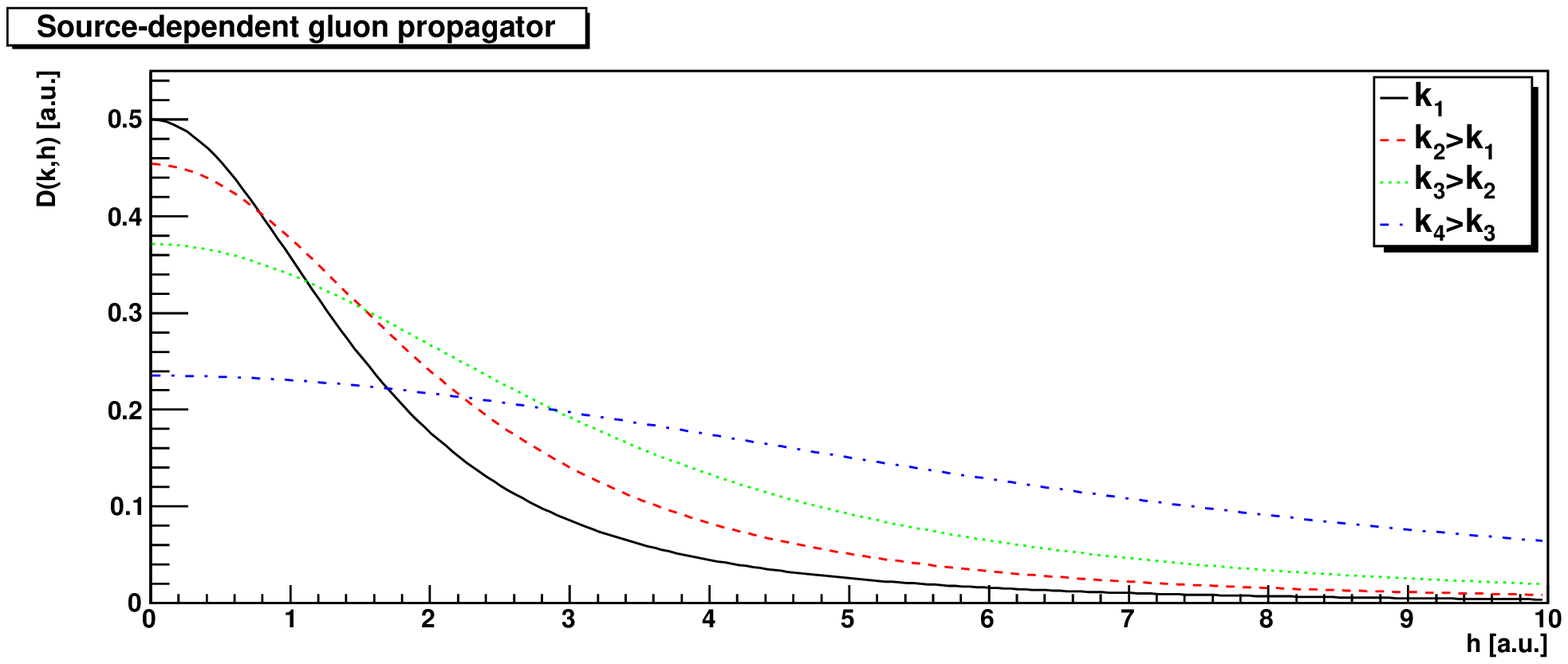}\\
\includegraphics[width=\linewidth]{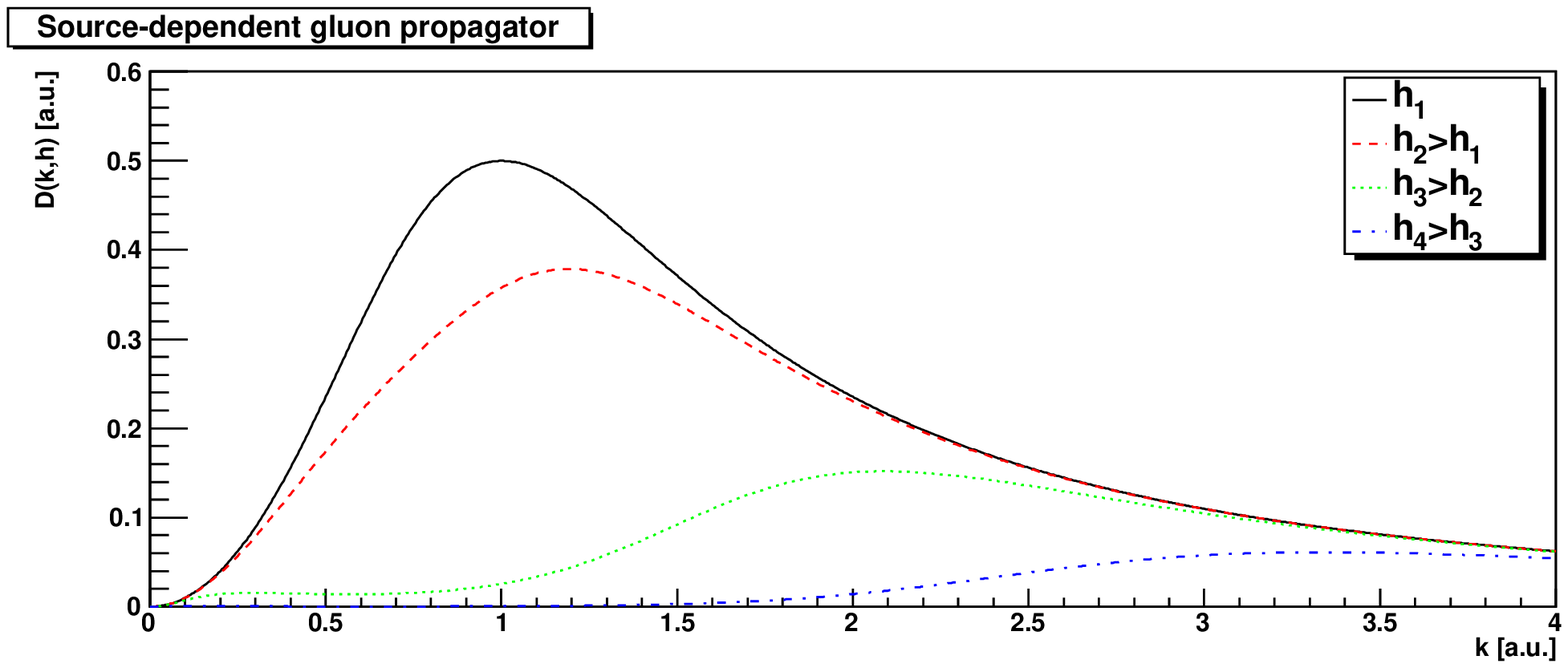}
\caption{\label{fig:d.gribov}The source-dependent Gribov-type gluon propagator, (\ref{gluonprop}), with $\hat\gamma = 1 + k^4$, as a function of the source strength, for different momenta (top panel), and as a function of momentum for different source strengths (bottom panel).}
\end{figure}

We shall convert the bounds,
\beqa
w(0, h) & \leq & |h| (2\pi/L)
 \label{0bound}   \\
 w(k, h) & \leq & |h k|/ \sqrt 2 \ {\rm for} \ k >> 2\pi/L,
 \eeqa
established in section \ref{locatehorizon}, into bounds on the gluon propagator $D(k, h)$ in the presence of the source $h$.  We have
\beq
D(k, h) = 2 {\p^2 w(k, h) \over \p h^2} = 2 {\p m(k, h) \over \p h }, 
\eeq
where $m(k, h)$ is the ``magnetization" introduced in the last section, so
\beq
\int_0^\infty dh \ D(k, h) = 2 [ m(k, \infty) - 2 m(k, 0) ]
\eeq
or, by (\ref{boundmag}),
\beq
\label{intgluprop}
\int_0^\infty dh \ D(k, h) \leq \sqrt 2 k,
\eeq
where the last inequality becomes an equality if the bound on $w(k, h)$ is saturated.  Recall that $D(k, h) \geq 0$ is positive, so the left-hand side represents the area under the curve $D(k, h)$ at fixed $k$, and this area decreases toward $0$ as $k$ decreases, as illustrated in the top panels of figure \ref{fig:d} and \ref{fig:d.gribov}.

This bound holds for $k >> 2 \pi / L$, and we take the infinite-volume limit, $L \to \infty$, keeping the momentum $k$ fixed.  In this case $k$ is any real number, and we take the limit $k \to 0$,
\beq
\lim_{k \to 0} \int_0^\infty dh \ D(k, h) = 0.
\eeq
Since $D(k, h) \geq 0$ is positive, this implies
\beq
\label{boundonprop}
D(k, h) = 0, \ {\rm \ for \ almost \ all} \ h \geq 0.
\eeq
In \cite{Zwanziger:1991} it was assumed that $w(0, h)$ and $D(0, h)$ are analytic in $h$, and it was concluded that this bound, which holds for {\em almost} all $h \geq 0$, does hold,  $D(0, h) = 0$, for {\em all} $h \geq 0$, giving a vanishing gluon propagator at $k = h = 0$.  However, numerical data, taken at $h = 0$, indicate that the gluon propagator is positive at 0 momentum, $\lim_{k \to 0} D(k, 0) > 0$, in Euclidean dimension $d = 3, 4$,  \cite{Cucchieri:2007rg, Cucchieri:2007md, Cucchieri:2010xr, Bogolubsky:2007ud, Bogolubsky:2009dc, Bornyakov:2009ug}, while $\lim_{k \to 0} D(k, 0) = 0$, for $d = 2$ \cite{Maas:2007uv, Cucchieri:2007rg, Cucchieri:2011ig}.\footnote{The vanishing of $D(k = 0)$ in Euclidean dimension $d = 2$ is proven in \cite{Zwanziger:2012,Cucchieri:2012cb,Huber:2012zj}.}  If so, we must allow for the possibility of a discontinuity of $D(0, h)$ at $h = 0$.  Ignoring the possibility of other discontinuities, we conclude from the last bound,\footnote{In the same way, for the zero momentum component, $k = 0$, on a finite volume $L^d$, we can show,
\beq
\int_0^\infty dh \ D(0, h) \ \leq  2 \pi / L.
\nonumber
\eeq
In the infinite-volume limit, we get
\beq
\lim_{L \to \infty} D(0, h) = 0.
\nonumber
\eeq
and
\beq
\lim_{h \to 0} \lim_{L \to \infty} D(k, h) = 0.
\eeq}
\beq
\label{propvanishes}
\lim_{h \to 0} \lim_{k \to 0} D(k, h) = 0.
\eeq
Thus if we first take the limit $k \to 0$, followed by $h \to 0$, then the gluon propagator vanishes at $k = 0$.

The numerical result at $H = 0$ for $d = 3, 4$ may be expressed as $\lim_{k \to 0}\lim_{h \to 0} D(k, h) > 0$.  Thus, by comparison with our proven result (\ref{propvanishes}), the lattice data in Euclidean dimension $d = 3, 4$ indicate that the order of limits does not commute.  We next exhibit a simple model in which the limits do not commute.

\subsection{A simple model free energy}\label{simplemodel}

As will be seen below, the numerical data are rather well described by the free energy per unit Euclidean volume
\beq
w(k ,h)=\sqrt{\hat\gamma^2(k)+\alpha^2 h^2 k^2}-\hat\gamma(k),\label{wmodel}
\eeq
\noindent which is motivated by the results in section \ref{sqea}.  Here $\alpha$ is a constant that accounts for undersaturation of the bound $w(h, k) \leq |hk|$, but saturates the bound $w(h) \leq \alpha |hk|$, and $\hat\gamma(k)$ is a function of $k$ at our disposal.  The quantum effective action $\gamma(a)$ is given by the Legendre transformation
\beq
a(h) = {\p w(h) \over \p h} = { \alpha^2 h k^2 \over (\hat\gamma^2(k)+\alpha^2 h^2 k^2)^{1/2} },
\eeq
where $a$ represents the $k$-th Fourier component of the classical configuration,
\beqa
\gamma(a) & = & ha - w(h)
\nonumber  \\
& = & \hat\gamma(k) \Big[ 1 - \Big( 1 - {a^2 \over \alpha^2 k^2} \Big)^{1/2} \Big]. 
\eeqa
It is non-analytic at the Gribov horizon $a =  \pm \alpha k$, and is defined only in its interior $|a| \leq \alpha |k|$.

The corresponding gluon  propagator with source of strength $h$ is given by
\beq
\label{gluonprop}
D(k,h) = {\p^2 w(k, h) \over \p h^2}=\frac{\alpha^2 k^2 \hat\gamma^2(k)}{(\hat\gamma(k)^2 + \alpha^2 h^2 k^2)^\frac{3}{2}}.
\eeq
With $w(k,h) \approx \alpha |h k|$ at large $h$, the bound (\ref{intgluprop}) on the gluon propagator reads $\int d^d h \ D(k, h) \leq \alpha k$, and we find
\beq
\int d^d h \ D(k, h) = \alpha k,
\eeq
independent of $\hat\gamma(k)$. 

In ordinary lattice calculations, the value $h = 0$ is set from the start, and for this value the model yields
\beq
D(k,0)=\frac{\alpha^2 k^2}{\hat\gamma(k)}.
\eeq
\noindent Since $\hat\gamma(k)$ is arbitrary, we may choose it to obtain any gluon propagator $D(k, 0)$ at $h = 0$ one wishes, and the model gives the $h$-dependence for all $h$.  As examples, in figure \ref{fig:d} $\hat\gamma(k)$ is chosen to produce the propagator $1/(k^2 + m^2)$, whereas in figure \ref{fig:d.gribov} it is chosen to produce the Gribov propagator $k^2/(k^4 + m^4)$.  In the first case the limits do not commute, and in the second they do.

The choice $\hat\gamma(k)=\alpha^2 k^2m^2$ gives a finite zero-momentum gluon propagator
\beq
\label{oneorder}
\lim_{k \to 0} \lim_{h \to 0} D(k,h) = \lim_{k \to 0}D(k,0)= { 1 \over m^2},
\eeq
as is observed in lattice calculations in dimension $d = 3$ or 4.  On the other hand, for this choice of $\hat\gamma(k)$, if we first take $k$ to zero, with $h > 0$, we find
\be
\lim_{k\to 0}D(k,h) = \lim_{k\to 0} {\alpha^3 k^3 m^4 \over h^3} = 0,
\eeq
for all $h > 0$, which gives
\beq
\label{anotherorder}
\lim_{h \to 0} \lim_{k \to 0} D(k, h) = 0.
\eeq
This agrees with the bound (\ref{boundonprop}), but disagrees with the order of limits (\ref{oneorder}).  Thus, this simple model reproduces exactly the results of numerical studies at $h = 0$ for which the gluon propagator is finite at $k = 0$, while satisfying the exact bound $\lim_{k \to 0} D(k, h) = 0$ for all $h > 0$.  This hinges critically on the non-analyticity of the model free energy (\ref{wmodel}), for which the radius of convergence in $h$ is $\hat\gamma(k)/\alpha k$, which vanishes with $k$ for $\hat\gamma(k) = \alpha^2 k^2 m^2$.  In this case, $w(k, h)$ becomes non-analytic in $h$ in the limit $k \to 0$.  Note however that this does not necessarily imply that the gluon propagator must be finite if $h$ is taken to zero first --- that depends on the actual form of $\hat\gamma(k)$. So, for example, if we take $\hat\gamma(k) = \alpha^2 m^4$, independent of $k$, we get, for $k$ small, $D(k, 0) \approx k^2/m^4$, which is the Gribov form, and in this case, the order of limits commutes.  

It is time to turn to the full theory to see what happens there.

\section{Numerical study}\label{snumeric}

\subsection{General considerations and systematic errors}

To test the predictions would, in principle, require the numerical measurement using a Monte Carlo approach of a current-dependent free energy, which can be split as
\begin{eqnarray}
\exp(W\left(J\right))&=&\int {\cal D} A \exp\left(-S\left(J\right)\right)\label{wdef}\\
S\left(J\right)&=&S\left(0\right)-\int JA\label{sj}\\
S\left(0\right)&=&\int\frac{1}{4}F_{\mu\nu} F_{\mu\nu} - \ln \det [M(A)]\label{s0}\\
\Delta S\left(J\right)&=&S\left(J\right)-S\left(0\right)\label{ds}
\end{eqnarray}
\noindent with the external current $J$. Since this current is gauge-dependent, any numerical update would be required to remain within the same gauge. But there is no yet an efficient lattice update algorithm known, which keeps the gauge fixed.

To circumvent this problem, reweighting will be used here. In this case, instead of creating a Markov chain based on $S\left(J\right)$, it will be created using $S\left(0\right)$. Then, $W$ will be obtained by the measurement
\begin{equation}
\exp(W\left(J\right))=\left\langle\exp\left(\int JA\right)\right\rangle,\label{wsubstitute}
\end{equation}
\noindent which will be performed in a fixed gauge, the minimal Landau gauge \cite{Maas:2011se}. If the quantity measured were a polynomial in the fields, only the usual caveats of lattice calculations would be required \cite{Montvay:1994cy}. However, it is exponential in the field. Thus, the importance sampling of standard update algorithms cannot be expected to be accurate, especially if $\Delta S\gtrsim S\left(0\right)$, i.\ e.\ when the exponential weight of the source term becomes comparable to the action itself. Of course, it cannot be excluded that already a small $\Delta S$ upsets the importance sampling significantly.

Concentrating on the example source (\ref{source})
\begin{equation}
J_\mu^a=\delta_{a3}\delta_{\mu 2} h\cos(x_1 k)\nonumber,
\end{equation}
\noindent which will be used in the following with $h$ always positive, a better estimate can be made. Because the gauge field in lattice units is bounded by one, the source term is of maximum size
\begin{equation}
\max\Delta S=\max\int dx JA=Vh\label{jbound}.
\end{equation}
\noindent The conventional action term is of typical size
\begin{equation}
S\left(0\right)=\frac{\beta V d(d-1)}{2}\langle P\rangle\nonumber,
\end{equation}
\noindent where $d$ is the number of space-time dimensions, $\beta=4/g^2$, and $\langle P\rangle$ is the plaquette expectation value, i.\ e.\ the free energy per unit volume. Thus, the maximum $h$ possible is expected to be of order
\begin{equation}
h\lesssim \frac{\beta d(d-1)\langle P\rangle}{2}\label{hmax}.
\end{equation}
\noindent If, as discussed above, the source term were bounded by
\begin{equation}
\max\Delta S=\frac{2\pi}{L}h\nonumber,
\end{equation}
\noindent instead of (\ref{jbound}), the situation would improve, as then not only for finer lattices but also for larger lattices the maximum possible value of $h$ would increase. But since this lower limit holds only for special field configurations, this may not be reliable, and in the following the more conservative estimate (\ref{hmax}) will be used.

\subsection{Lattice setup}

Due to the reweighting approach, the numerical simulation can be performed using standard methods. Using the Wilson action, configurations were created using a hybrid heat-bath-overrelaxation update and then gauge-fixed to minimal Landau gauge using stochastic overrelaxation, see \cite{Cucchieri:2006tf} for details.

\begin{table}
 \begin{tabular}{|c|c|c|c|c|c|}
 \hline
     & $a$ [fm] & 0.20 & 0.10 & 0.05 \cr
\hline
\hline
  $d=2$ & $\beta$ & 7.99 & 30.5 & 120 \cr
\hline
   & $L=60$ & 1938 & 1720 & 2128 \cr
  & $V$ & $(12$ fm$)^2$ & $(6$ fm$)^2$ & $(3$ fm$)^2$ \cr
  & $L=120$ & 1034 & 1066 & 2160 \cr
  & $V$ & $(24$ fm$)^2$ & $(12$ fm$)^2$ & $(6$ fm$)^2$ \cr
  & $L=240$ & 330 & 1560 & 590 \cr
  & $V$ & $(48$ fm$)^2$ & $(24$ fm$)^2$ & $(12$ fm$)^2$ \cr
\hline
\hline
  $d=3$ & $\beta$ & 3.73 & 6.72 & 12.7 \cr
\hline
  & $L=18$ & 2277 & 2277 & 3415 \cr
  & $V$ & $(3.6$ fm$)^3$ & $(1.8$ fm$)^3$ & $(0.9$ fm$)^3$ \cr
  & $L=36$ & 2261 & 2258 & 3356 \cr
  & $V$ & $(7.2$ fm$)^3$ & $(3.6$ fm$)^3$ & $(1.8$ fm$)^3$ \cr
  & $L=72$ & 2228 & 2296 & 4520 \cr
  & $V$ & $(14.4$ fm$)^3$ & $(7.2$ fm$)^3$ & $(3.6$ fm$)^3$ \cr
\hline
\hline
  $d=4$ & $\beta$ & 2.221 & 2.457 & 2.656 \cr
  & $L=6$ & 3202 & 2406 & 1050 \cr
  & $V$ & $(1.2$ fm$)^4$ & $(0.6$ fm$)^4$ & $(0.3$ fm$)^4$ \cr
  & $L=12$ & 1792 & 1012 & 1840 \cr
  & $V$ & $(2.4$ fm$)^4$ & $(1.2$ fm$)^4$ & $(0.6$ fm$)^4$ \cr
  & $L=24$ & 1149 & 2296 & 1176 \cr
  & $V$ & $(4.8$ fm$)^4$ & $(2.4$ fm$)^4$ & $(1.2$ fm$)^4$ \cr
\hline
 \end{tabular}
 \caption{\label{configs}The considered physical systems and number of configurations. Scales and the lattice spacing $a$ as a function of the bare gauge coupling $\beta$ have been set using a string tension of $(440$ MeV$)^2$, according to \cite{Maas:2007uv,Cucchieri:2008qm,Cucchieri:2006tf}. The number of thermalization hybrid sweeps \cite{Cucchieri:2006tf} is $2(10L+100(d-1))$, where $L$ is the linear lattice size, and a tenth of this number is used for decorrelation between two consecutive measurements.}
\end{table}

Since, as noted, two dimensions is found to behave rather differently on the level of the propagators than higher dimensions, here the free energy (\ref{wdef}) will be determined for $d=2$, 3, and 4. To obtain an estimate of both finite volume and finite lattice spacing artifacts, in all dimensions nine different lattice settings have been investigate, see table \ref{configs} for details.

After gauge-fixing, the gluon fields are determined in a standard way from the links \cite{Maas:2011se,Cucchieri:2006tf}. The determination of (\ref{wsubstitute}) with the source (\ref{source}) is then straightforward. However, for large values of $h$ standard long double precision is insufficient due to the exponential behavior, and arbitrary precision arithmetic was necessary, especially for a reliable determination of the statistical errors. However, in a small window close to the point when switching between fixed and floating precision, the required number of digits became large, and as a consequence the statistical errors in this region are somewhat overestimated.

Furthermore, especially at small $h$, it can happen that the exponent of (\ref{wsubstitute}) fluctuates statistically to values smaller than zero, thus giving an average of the exponential smaller than one, and consequently $W$ becomes negative. This is a purely statistical artifact, and only occurs within statistical errors. In the limit of infinite statistics, $W$ is indeed positive and convex. However, this seriously affects the statistical reliability of $W$ at small $h$, while the opposite effect occurs at large values of $h$.

Below, also the gluon propagator at zero momentum will be used. It is again determined using standard methods, see \cite{Maas:2011se,Cucchieri:2006tf} for details.

\subsection{Checking the systematic limit}

\begin{figure}
\includegraphics[height=0.8\textheight]{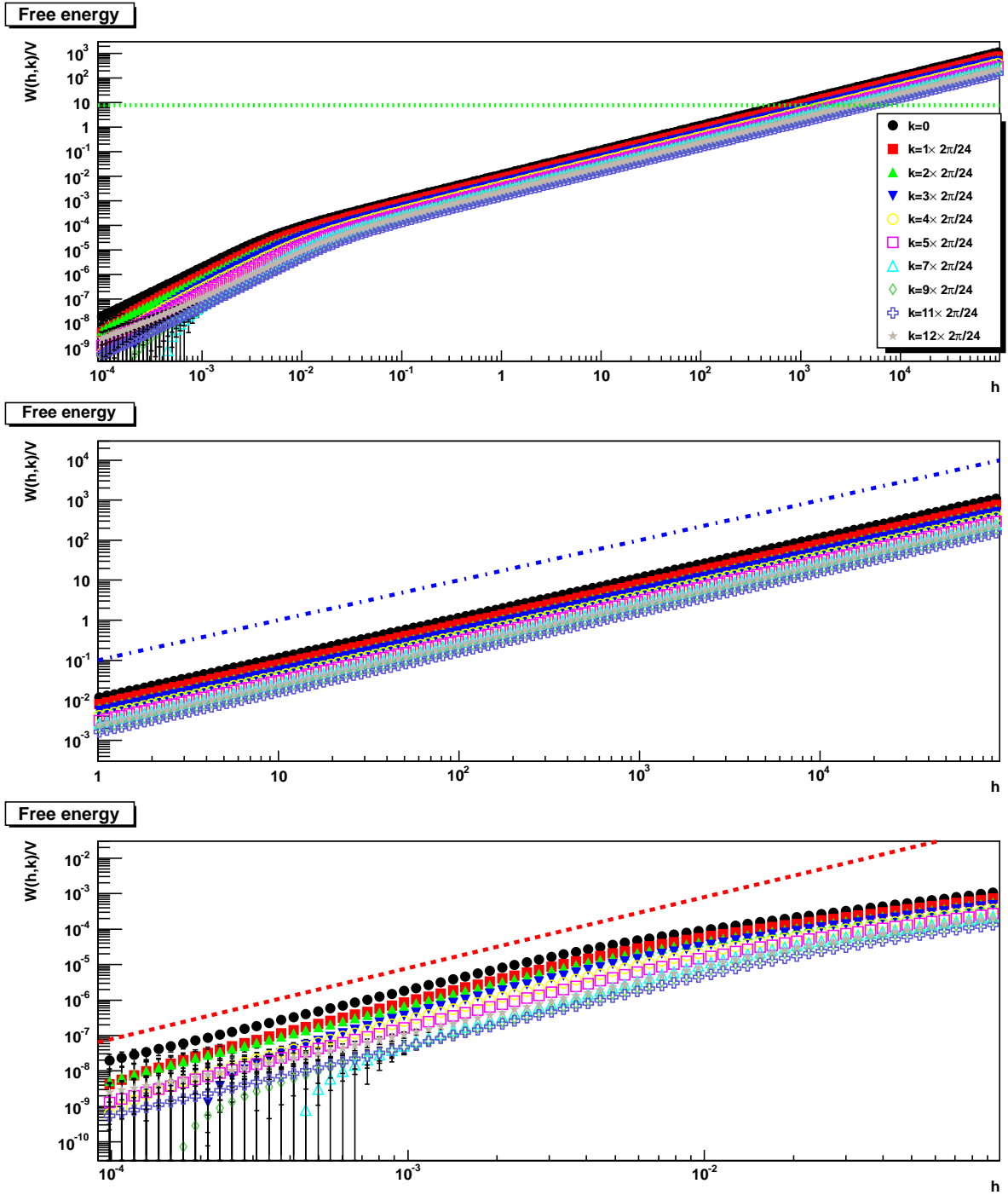}\\
\caption{\label{wres}The free energy density $W(h,k)/V$ in lattice units as a function of $h$ for various vales of $k$. The middle and bottom panels show magnifications of the region at large and small $h$, respectively. The value of $h$ is limited by (\ref{hmax}), indicated by the green dashed line in the top panel. In the middle panel the blue dashed line is linear in $h$, while the red dashed line the bottom panel is proportional to $h^2$. The four-dimensional $24^4$ lattice has at $\beta=2.221/a=0.2$ fm a physical volume of (4.8 fm)$^4$.}
\end{figure}

With the source (\ref{source}), the functional $W$ becomes a function of the two independent variables $k$ and $h$. Since $k$ is a lattice momentum, it can have only discrete values, while $h$ is a continuous variable. For $k$, at most ten different values have been used, depending on the lattice size. An example for $W$ is shown in figure \ref{wres}. The situation for all other lattice settings is virtually indistinguishable; without labeling, the plots cannot be separated from one another. It is immediately visible that $W$ depends linearly on $h$ at large values of $h$, and quadratically at small values of $h$. Since this sets in already some orders of magnitude below the reliability limit (\ref{hmax}), this may be indeed a genuine effect. Based on the argumentation in section \ref{sanamc}, we expect that to be the genuine behavior, which is thus quadratic in $h$ at small $h$ and linear in $h$ at large $h$, manifesting the behavior predicted in section \ref{soptbound}.

It is an interesting observation that the error appears to become smaller at large $h$, which is at first sight contradictory to the expectations for reweighting. To understand this, it should be recalled that actually not $W$ is measured, but rather $\exp W$, and a logarithm is taken, and the statistical error is propagated. Since, at small $h$, $\exp W$ is exponentially close to one, the error is exponentially increased when taking the logarithm. At the same time, at large $h$, $W$ itself is large, and the error on it is exponentially suppressed. This is only due to statistical fluctuations. The systematic error due to reweighting is not captured.

\subsection{Free energy and asymptotic behavior}

\begin{figure}[h]
\includegraphics[width=0.333\linewidth]{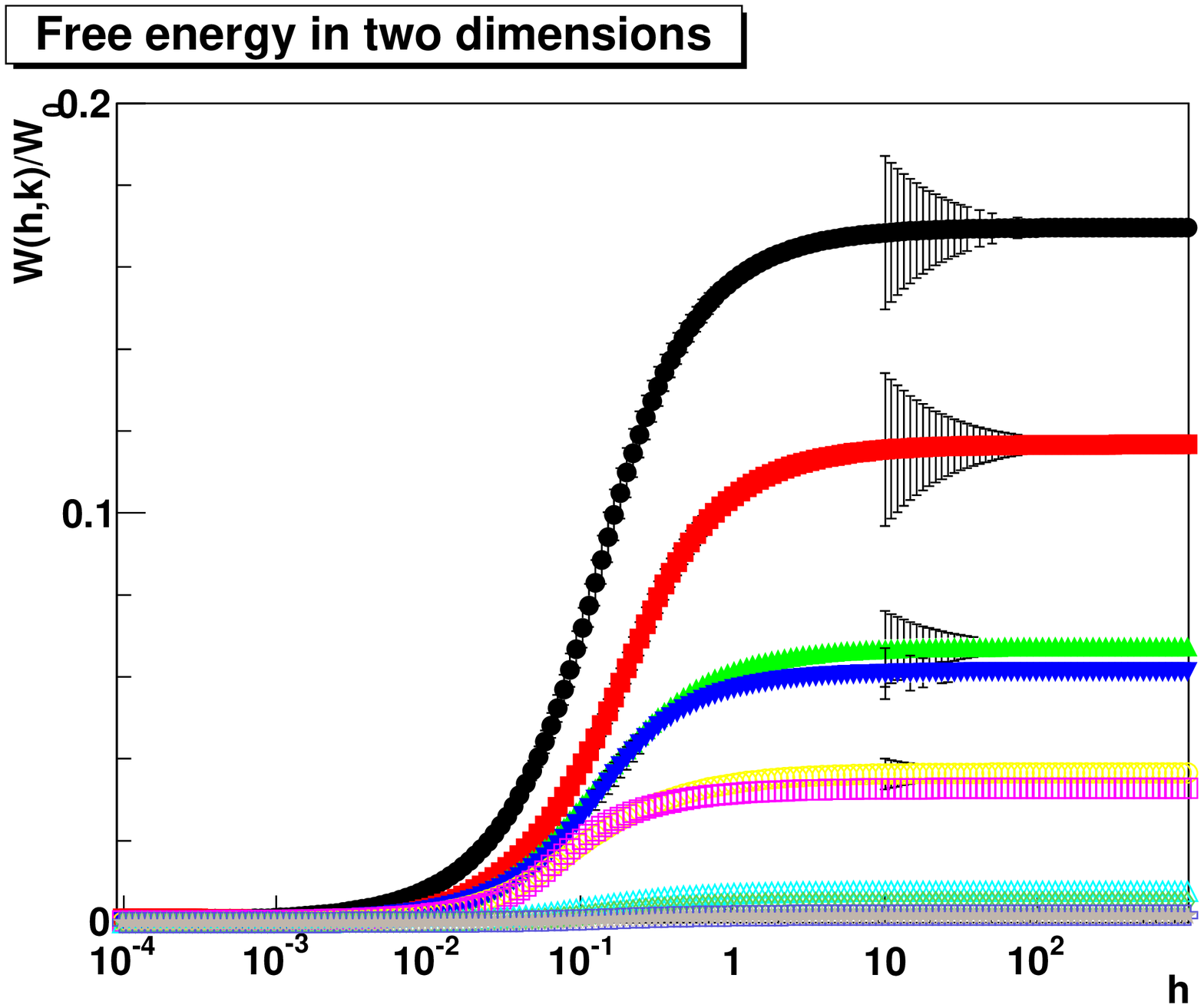}\includegraphics[width=0.333\linewidth]{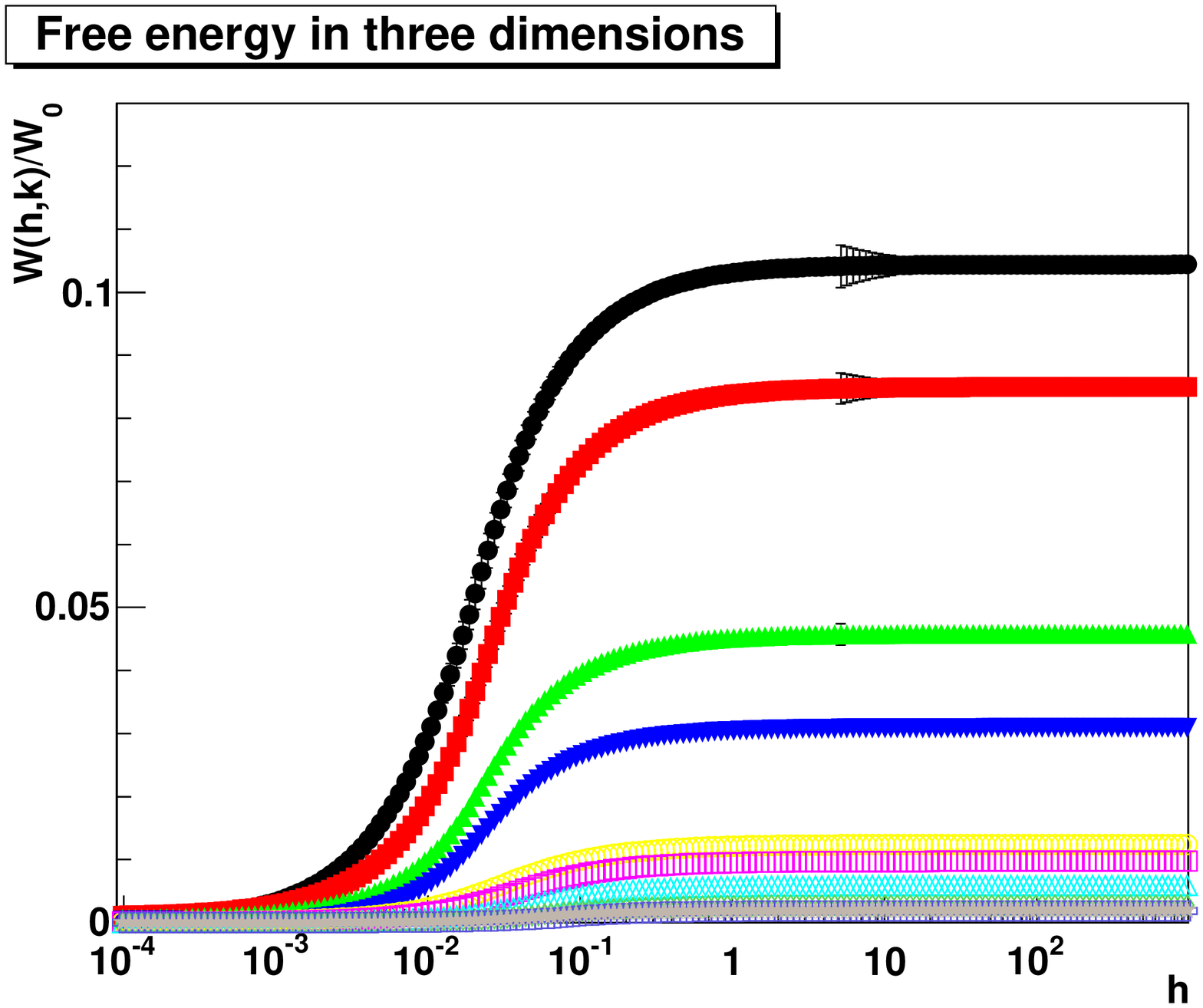}\includegraphics[width=0.333\linewidth]{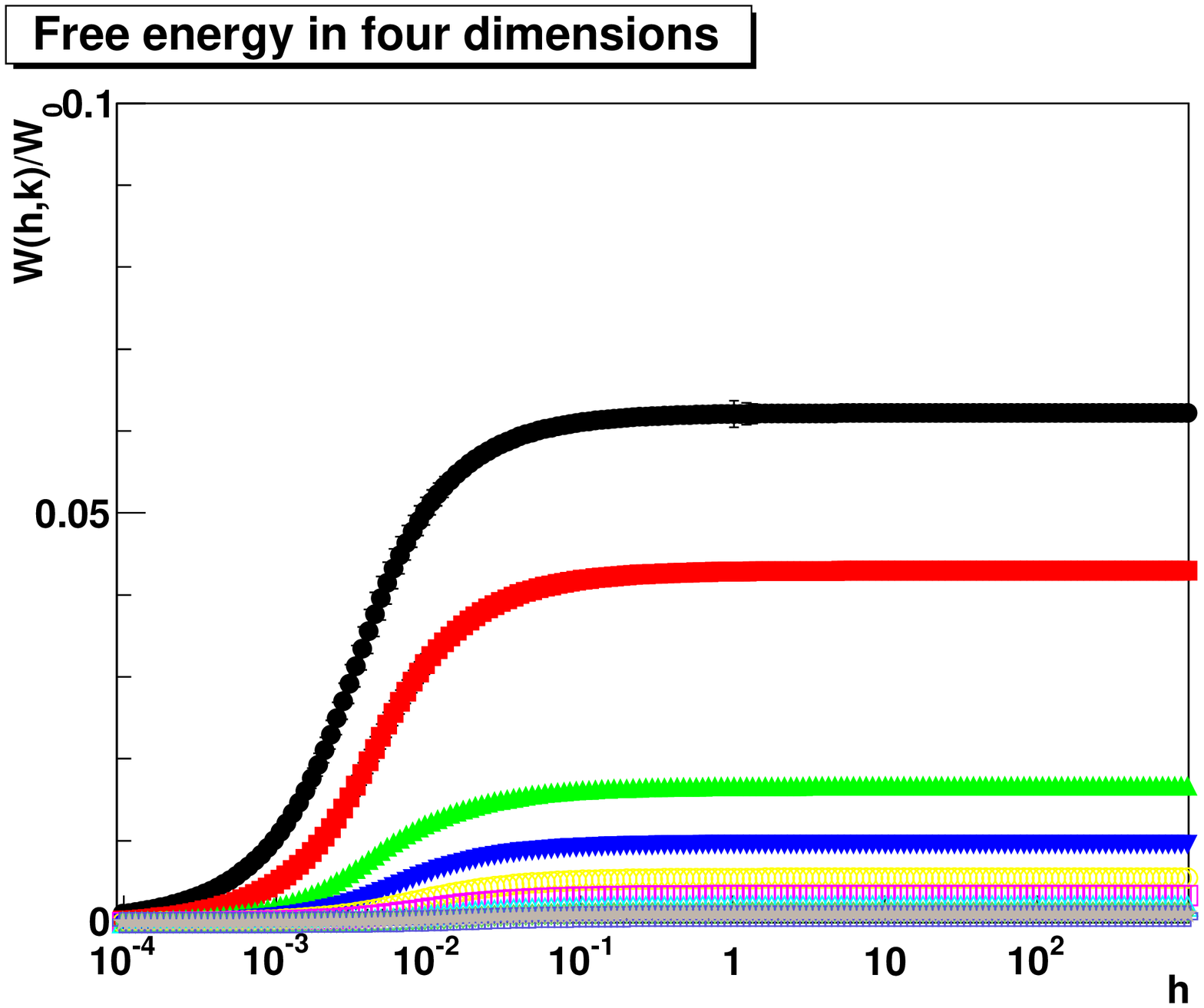}
\caption{\label{wnorm}The normalized free energy $W/W_0$ as a function of $h$ for two (left panel, 120$^2$/(24 fm)$^2$ at $\beta=7.99$/$a=0.2$ fm), three (middle panel, 36$^3$/(7.2 fm)$^3$ at $\beta=3.73$/$a=0.2$ fm), and four (right panel, 24$^4$/(4.8 fm)$^4$ at $\beta=2.221$/$a=0.2$ fm) dimensions. The momentum increases in all cases from top to bottom. The intermediate increase in the statistical error is due to the aforementioned switch to fixed precision numerics.}
\end{figure}

To identify the slope more precisely, figure \ref{wnorm} shows $W/W_0$, where $W_0$ is $\sqrt{2}\pi L^{d-1} h$ for $k=0$ and $L^d/\sqrt{2} h k$ otherwise, motivated by the limits in section \ref{locatehorizon}. The free energy indeed shows the expected large $h$ behavior, and any correction to it is smaller than $\ln(h)$, as has been explicitly tested. However, the bound $W_0$ is not saturated, and a prefactor smaller than one remains. Based on the arguments in section \ref{sanamc}, this was to be expected.

\begin{figure*}
\includegraphics[width=\linewidth]{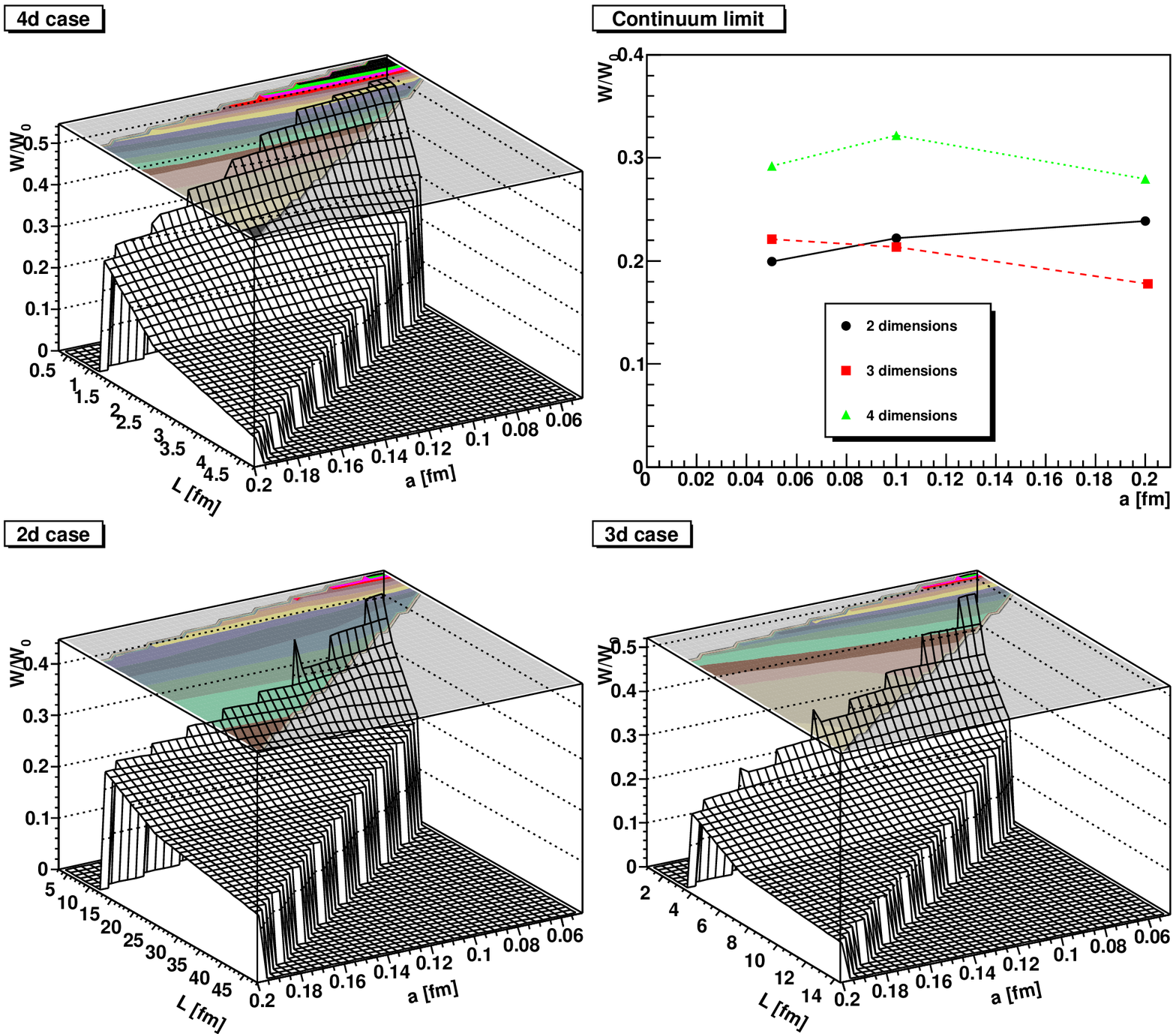}\\
\caption{\label{limits}The value of the normalized free energy density $W(h,k)/W_0$ at $k=0$ in the domain at large $h$, where it becomes constant, as a function of lattice volume and discretization for two (bottom left panel), three (bottom right panel) and four (top left panel) dimensions. The top right panel shows a cut at fixed spatial volume. Statistical error bars are smaller than the symbols. The volume is (12 fm)$^2$ in two dimensions, (3.6 fm)$^3$ in three dimensions, and (1.2 fm)$^4$ in four dimensions.}
\end{figure*}

To investigate whether this undersaturation is a lattice artifact or depends on the dimensionality, the remaining constants for all systems of table \ref{configs} have been determined, and are shown in figure \ref{limits}. At first sight, no qualitative difference is found. It is visible that at fixed lattice spacing the expected bound is less fulfilled the larger the volume. No final conclusion can be drawn from this, except that the order of limits may be important, and that an undersaturation remains in all cases investigated here, and that this undersaturation seems to not decrease towards the continuum and thermodynamic limit.

\subsection{Quantum effective action}

Once the free energy is known, it is a straight-forward exercise to also determine its Legendre transform (\ref{legendre}) numerically, the quantum effective action. The genuine advantage is that the Legendre transform shows different properties, and thus a fit which describes both the free energy and the quantum effective action correctly will certainly capture more of the pertinent features than a fit which describes just one.

Especially, the discussion and arguments presented in section \ref{sqea} and \ref{sgp} suggest strongly a fit form of type
\beqa
W(h,k)&=&Vc\left(\sqrt{1+m^{-2d} h^2 k^2} - 1\right),\label{fitform}\\
W(h,0)&=&Vc\left(\sqrt{1+m^{-2d + 2} h^2} - 1\right),\nonumber
\eeqa
\noindent where $c$ and $m$ are fit parameters. This fit form has the expected asymptotic dependencies on $h$. The classical field $A$ is then defined as the derivative of $w=W/V$ w.\ r.\ t.\ the source strength $h$, and after resolving the implicit dependence the Legendre transform yields the quantum effective action. Note that the $k$-dependent function $\hat\gamma$ of (\ref{gluonprop}) corresponds here to a suitable combination of the fit parameters $c$ and $m$, which therefore are fitted for every $k$ independently.

\begin{figure*}
\includegraphics[width=\linewidth]{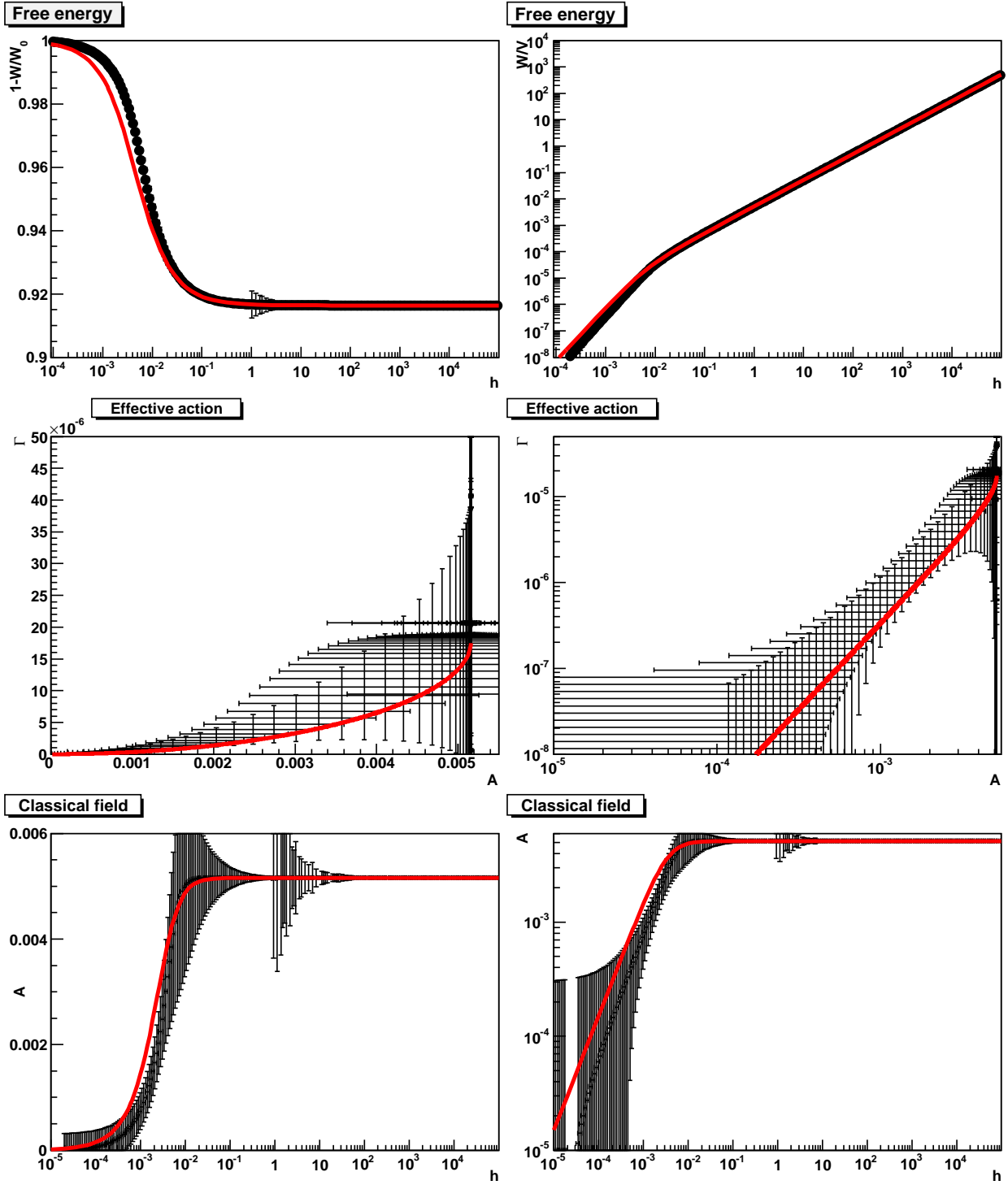}\\
\caption{\label{fig:gamma}The free energy (top panels), quantum effective action (middle panels), and classical field (bottom) compared to the fit function (\ref{fitform}) (thick line). Results are for a $(14$ fm$)^3$ lattice with $a=0.2$ fm.}
\end{figure*}

This can be applied both on the numerical data, using numerical derivatives, as well as on the analytical fit. To compare the two, the fit parameters are determined from the free energy. The result for a sample lattice setting at $k=0$ is shown in figure \ref{fig:gamma}. Though the statistical errors become large due to the numerical derivatives, it is clearly visible that the fit form describes the free energy, the classical field, and the quantum effective action acceptable. Especially, it is nicely seen how the Gribov horizon manifests itself in the form of a divergence of the quantum effective action at a finite classical field value, like in figure \ref{fig:gamma-illu}, and that the classical field is indeed bounded. It should be noted that this value of the classical field is indeed bounded by a value very small compared to the maximum field of one in lattice units. The Gribov horizon is thus small compared to the maximal fluctuations permitted for the gluon field, and cut-off effects should thus be small.

The situation for the other lattice settings is essentially the same. The fits yield always a mass parameter $m$ of typical size a few hundred MeV, though significantly dependent on lattice spacing, discretization, and dimensionality, with the tendency to rise with the dimension.

Thus, the form (\ref{fitform}) is a remarkably good description of the free energy of Yang-Mills theory, giving support to the arguments given in section \ref{sgp} that indeed a non-analyticity is present.

\subsection{Source-dependent propagator}

With respect to the question of how the free energy and the propagators can be related, there are two critical tests, which can be made. One is, whether the free energy is analytic in the source strength parameter in the infinite-volume limit. The other one is whether the source-dependent propagators approach their source-independent ones in the limit of $h$ in the infinite-volume limit. As discussed in section \ref{simplemodel}, this cannot be expected for a simple model. Since this simple model is a surprising good fit of both the free energy and the quantum effective action, this may be suspected also for the full case.

The first test is therefore the analyticity. This can be tested by checking whether at small $h$ the free energy is well approximated by the first terms of its Taylor series. Because $W(h,k)$ is the generating functional of connected correlation functions, see (\ref{genfunc}),  it should therefore be given to leading order by the gluon propagator at zero external field $D(k,0)$, as\footnote{Note that here and hereafter the convention $\theta(0)=1/2$ is used.}
\be
W(k, h)\approx\frac{V}{4\theta(k)}D(k,0)h^2+{\cal O}(h^4)=W_l(k,h)+{\cal O}(h^4)\label{wexpansion}.
\ee
\noindent Since this investigation is performed at small $h$ the reweighting issues should be less relevant for this analysis.

\begin{figure*}
\includegraphics[height=0.2\textheight]{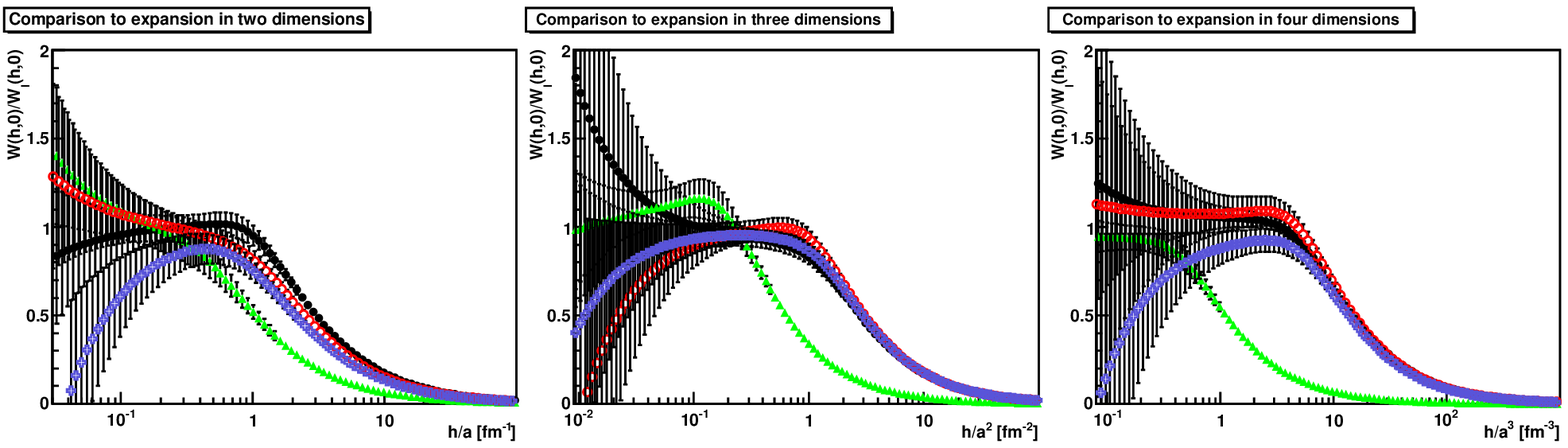}\\
\includegraphics[height=0.2\textheight]{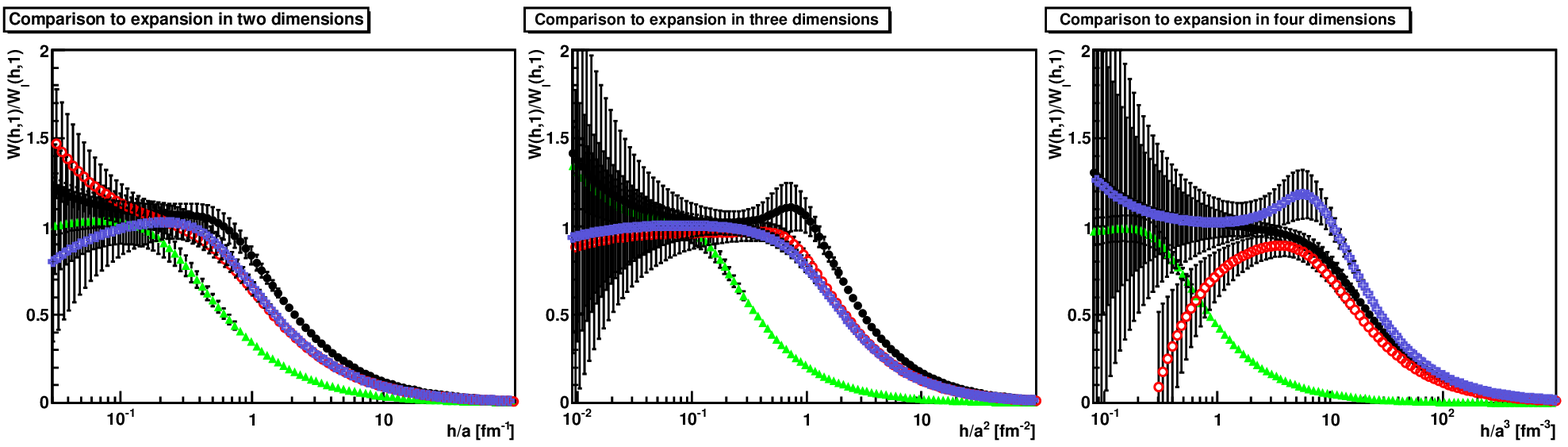}\\
\includegraphics[height=0.2\textheight]{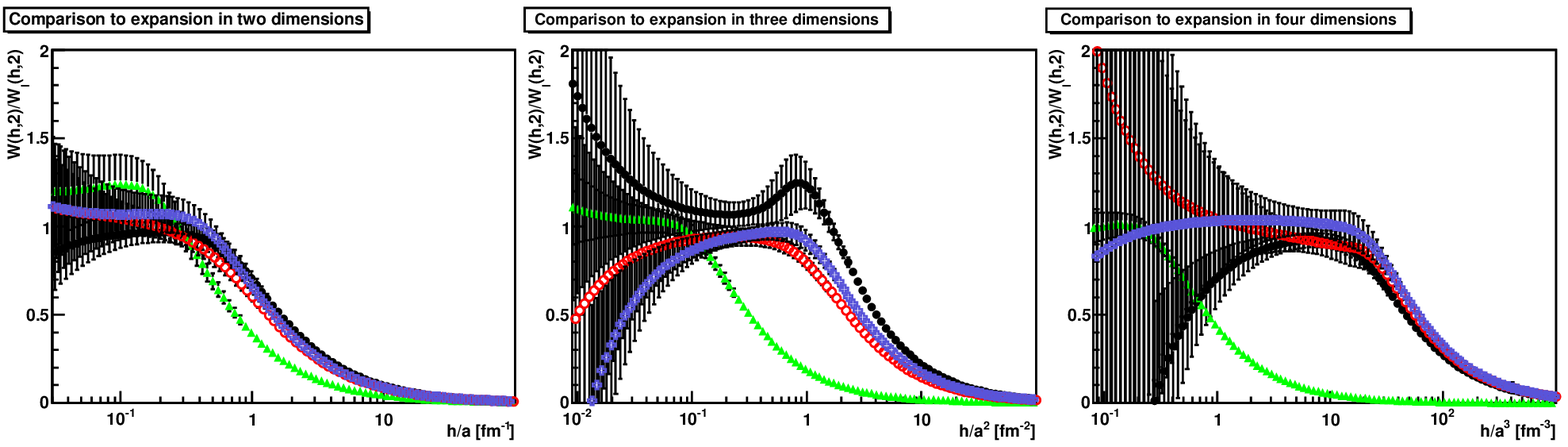}\\
\includegraphics[height=0.2\textheight]{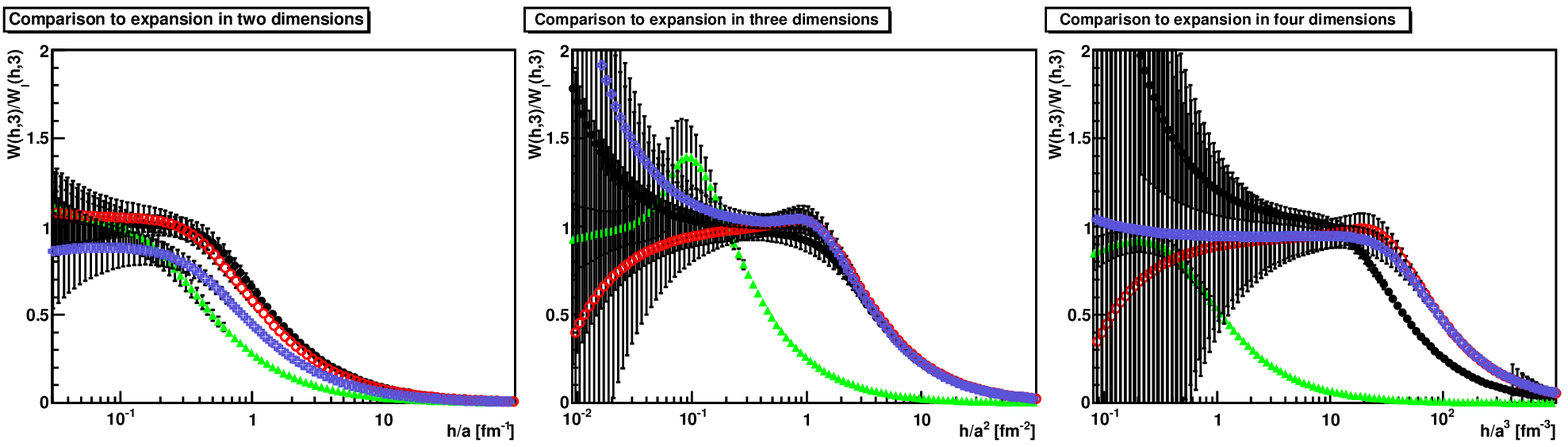}\\
\includegraphics[width=\textwidth]{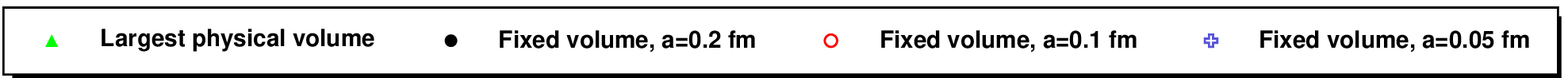}
\caption{\label{d0}The ratio of the free energy, divided by the leading Taylor coefficient as a function of the dimensionful field $h$. The fixed volume is (12 fm)$^2$ in two dimensions, (3.6 fm)$^3$ in three dimensions and (1.2 fm)$^4$ in four dimensions. The largest volume can be taken from table \ref{configs}. Results are shown for $2\pi L k=0,1,2,3$ from top to bottom.}
\end{figure*}

The comparison is made in figure \ref{d0} for various low momenta. Note that by taking the ratio any renormalization constants will drop out.

The result is interesting. First of all, in all cases the leading approximation is a good description at small $h$, at least within errors. However, with increasing volume this description becomes worse at fixed $h$ for all dimensions, though this effect is more pronounced the higher the dimension. At the same time, in four and three dimensions the continuum limit does not change the quality of the approximation at fixed $h$, while in two dimensions the approximation becomes marginally worse closer to the continuum limit. The approximation is further worsened when increasing the momentum. This is due to the presence of lattice spacing corrections. This is most notable for the largest momentum $k L / 2\pi = 3$, where for $a=0.2$ fm at $(1.3$ fm$)^4$ the approximation is worst. This is not surprising; in this case the lattice size was only $6^4$, and thus at $k=3$ the lattice structure is probed, leading to large corrections, which spoils the expansion (\ref{wexpansion}). In general, the lattice sizes are smallest in four dimensions, and thus the largest finite lattice spacing corrections would be expected there, which is also what is seen.

This result already suggest that the analyticity of the free energy is doubtful, though of course no numerical investigation can ever disprove it.

A second test is, whether the source-dependent gluon propagator $D(k,h)$ tends towards $D(k,0)$. The arguments in section \ref{sgp} suggest that this is not the case: The usual gluon propagator is not the limit of the source-dependent one.

To determine the source-dependent gluon propagator, there are two possibilities. One is to use the reweighting factor as a probability, yielding the positive semi-definite quantity
\beqa
D(k,h)&=&\frac{\partial^2 w(h)}{\partial h^2}=\frac{1}{V}\left(\left\langle\sum_x A_y^1(x) \cos(x k)\sum_y A_y^1(y) \cos(y k)\exp\left(\int JA\right)\right\rangle { 1 \over Z(J) }\right.\nonumber\\
&&-\left.\left\langle \sum_x A_y^1(x) \cos(x k)\exp\left(\int JA\right)\right\rangle^2 { 1 \over Z^2(J) } \right)\label{fdprop}.
\eeqa
\noindent Because the averages are taken with a selection of orbits obtained by the usual Boltzmann weight, this cannot expected to be accurate at very large $h$, once more an artifact from the reweighting. The second option is by a numerical derivative, i.\ e.\ using the formula
\be
D(k,h)=\frac{1}{2\theta(k)}\left.\frac{\partial^2 w}{\partial h\partial h}\right|_{k\text{ fixed}}\nonumber,
\ee
\noindent which is again only correct up to reweighting artifacts, and could be negative within statistical errors. As the formulation (\ref{fdprop}) includes also the first variation, both  are differently influenced by the statistical errors, which are determined from error propagation. Thus, in the numerical evaluation below for each value of $h$ and $k$ the formulation is used which has less statistical error. However, within the statistical errors both agree, though for large $h$ especially the statistical fluctuations of (\ref{fdprop}) become so large as to render this statement meaningless. Still, at small $h$ this supports that reweighting artifacts, which could affect both formulations differently, are not too large.

Note that this gluon propagator is anistropic in both momentum and color space, and here only the component along the source direction is regarded in both cases.

\begin{figure*}
\includegraphics[height=0.2\textheight]{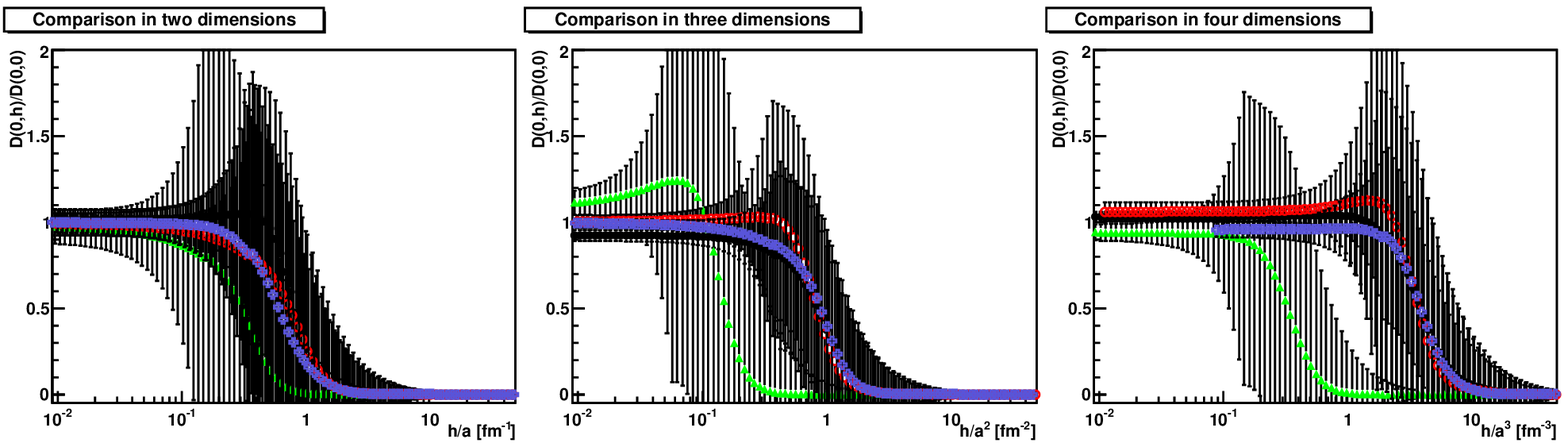}\\
\includegraphics[height=0.2\textheight]{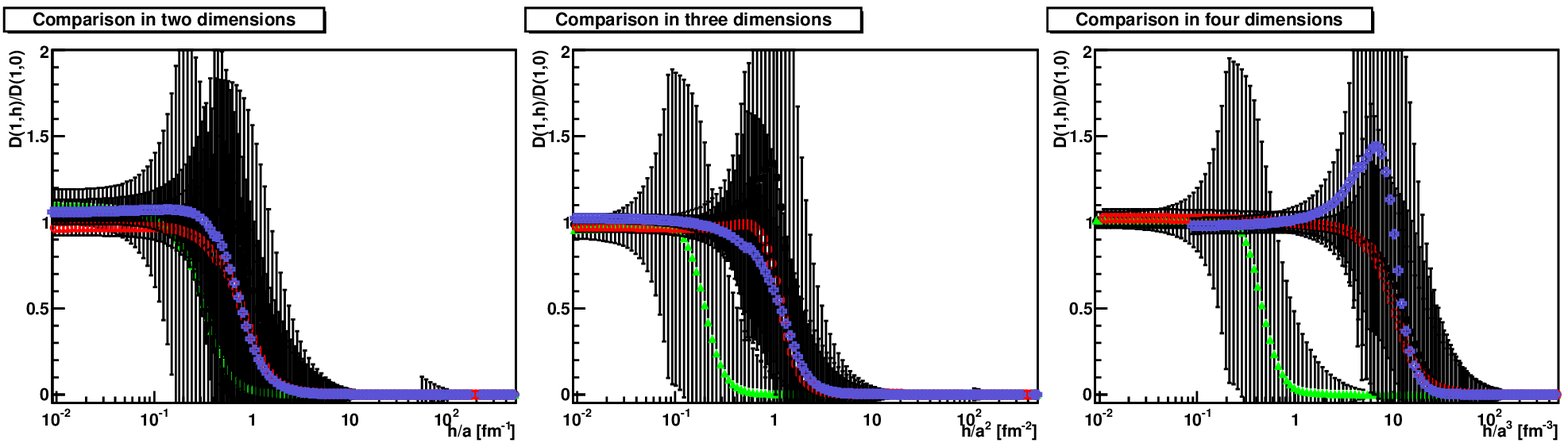}\\
\includegraphics[height=0.2\textheight]{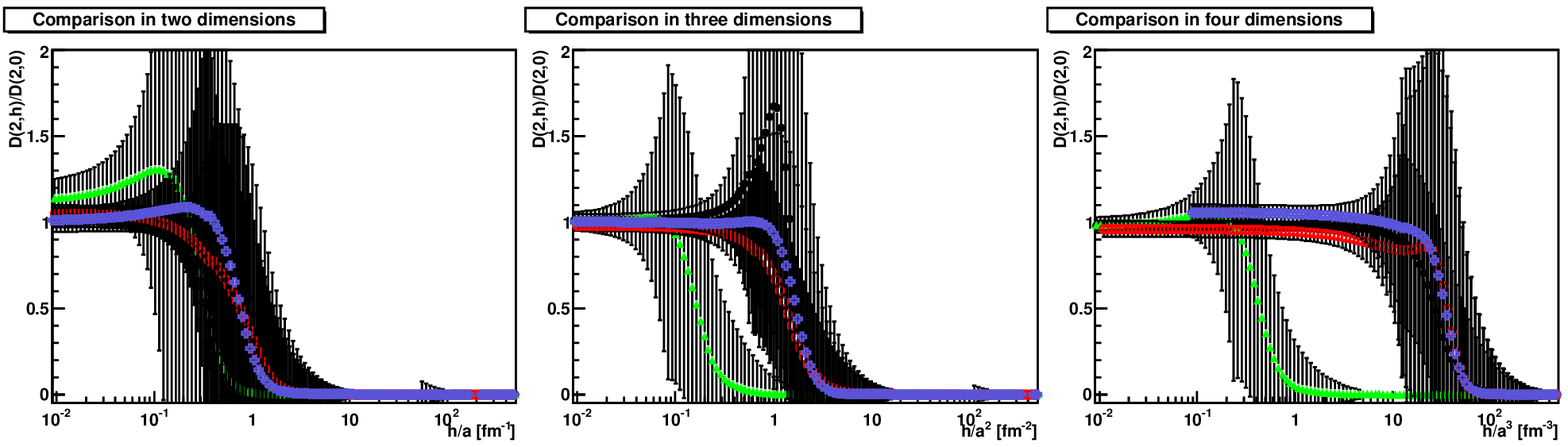}\\
\includegraphics[height=0.2\textheight]{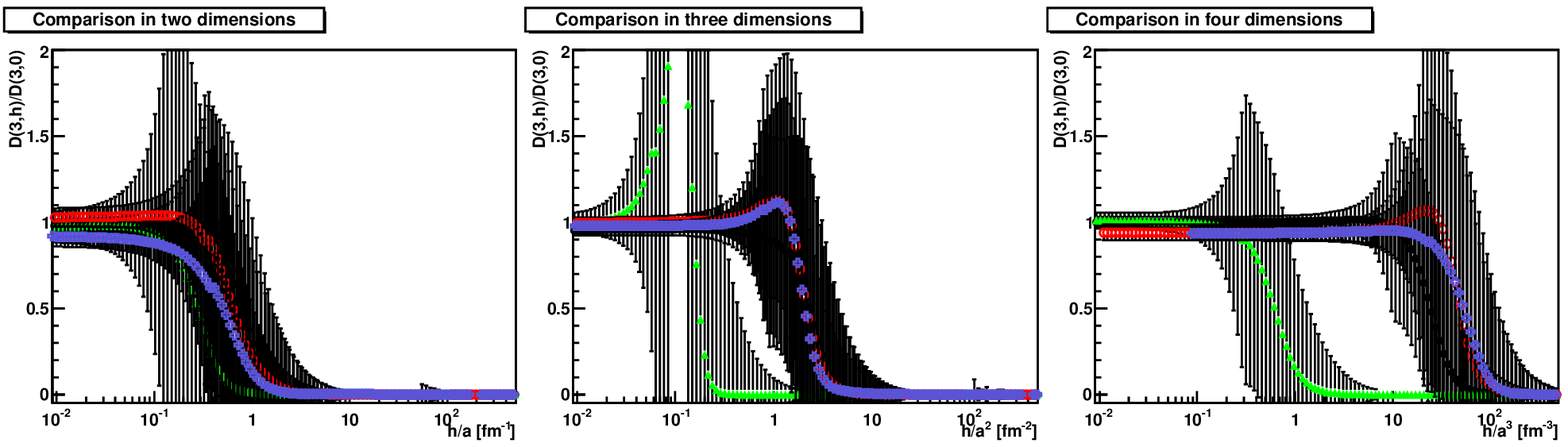}\\
\includegraphics[width=\textwidth]{legend.eps}
\caption{\label{dh0}The ratio of the source-dependent propagator and the source-independent propagator as a function of the source strength. The fixed volume is (12 fm)$^2$ in two dimensions, (3.6 fm)$^3$ in three dimensions and (1.2 fm)$^4$ in four dimensions. The largest volume can be taken from table \ref{configs}. Results are shown for $ k L / 2\pi =0,1,2,3$ from top to bottom.}
\end{figure*}

The result, normalized to the field-zero case, is shown in figure \ref{dh0}. There are a number of remarkable features.

First, at small source strength the ratio becomes one. This is to be expected, as there will be no non-analyticity in a finite volume, and therefore the source-dependent and source-independent propagator have to agree in the limit.

Second, while the lattice spacing effects are negligible, this is not the case for the volume-dependence. In fact, if the physical volume is increased, the ratio deviates at much smaller values of the source strength from one, and the earlier the higher the dimension. This is precisely what is expected if there is a non-analyticity, as argued in section \ref{sgp}.

Third, this effect sets in at larger source strength, the higher the momentum, i.\ e.\ the longer the wave-length the earlier the increase in size is felt.

Concluding, the numerical results are in favor of a non-analyticity in the free energy, and thus a non-equivalence of the propagators in the limit of zero source strength and at zero source strength. It should, however, not be forgotten that these results have been obtained using reweighting, and that they are available only on a limited range of lattice settings. Thus, this should not be taken as a final proof, but rather only as supporting evidence.

\section{Discussion and conclusion}

Assessing the implications of the presented findings is not simple. It is therefore best to first recapitulate some pertinent features.

What we set out to do was to understand why different approaches should yield different results for the same quantity, the gluon propagator. In the end, what we found is that the difference arises because it is not the same quantity.

To this end, one must reconsider the role of the free energy (or the quantum effective action) and its associated sources. In principle, the free energy, as the generating functional of correlation functions, appears to be an auxiliary mathematical construction, for correlation functions can be calculated by various methods, without introducing sources.  However sources can make this process very easy to formulate, and finite sources act as a control parameter to deform the physical system.  On the other hand, in gauge theories, external sources act in a gauge-dependent way, and therefore do not necessarily correspond to physical deformations. This does not, in principle, limit their usefulness; perturbation theory is an excellent example of this.  So while it is tempting to consider the deformation produced by gauge-dependent sources as somehow unphysical, nevertheless a mathematically valid formulation that may appear unphysical can nevertheless be very useful.  After all, imaginary time is unphysical, but Euclidean quantum field theory has proven its value.    

The question arises as to whether the source term $(J, A)$ provides a reweighting of different Gribov copies or a reweighting of gauge orbits or some of both.  We would like to emphasize that, at least to some extent, it is a reweighting of gauge orbits.  Indeed consider the case of a perfect gauge fixing so each gauge orbit has a unique representative.  This could be achieved, in principle, in the minimal Landau gauge if the absolute minimum on each orbit were chosen.  In this limiting case, the source term provides a pure reweighting of gauge orbits, which would become visible if a suitable gauge-invariant quantity were calculated as a function of the external source strength $h$, and some of this property presumably persists in the gauge fixing that we actually do.   However we have not tested this in the present article, but instead we have studied the free energy $W(k, h)$ and the gluon propagator, $D(k, h)$ as a function of momentum $k$ and source strength $h$.

Here, however, something interesting happens. The source we introduced is perfectly fine perturbatively. But in our non-perturbative calculation something changes.  We have proven that at infinite volume the gluon propagator $D(k, h)$ vanishes at zero momentum $k$ for every non-zero source strength, $\lim_{k = 0}D(k, h) = 0$, for all $h > 0$, {\em no matter how weak.}  Our lattice data are consistent with this behavior.  On the other hand, lattice studies done at $h = 0$ in $d$ = 3 and 4 dimensions \cite{Cucchieri:2007rg, Cucchieri:2007md, Cucchieri:2010xr, Bogolubsky:2007ud, Bogolubsky:2009dc, Bornyakov:2009ug}, including the present study,  give a finite value for the gluon propagator at zero momentum, $\lim_{k \to 0}D(k, 0) > 0$.  So, if one takes the lattice data at face value, there must be a jump in the low momentum limit of the gluon propagator at $h = 0$.  Stated differently, the value of the gluon propagator $D(k, h)$ depends on the order of limits at $k \to 0$ and $h \to 0$, and $D(k, h)$ is non-analytic in $h$ at $k = 0$.

In addition to the non-analyticity, we find that the gluon propagator is suppressed at all momenta, or even essentially vanishing for all momenta below a scale which is given by the external source strength. One could interpret this as a resistance of the system to an external field that tries to create color excitations; the system reacts by just shutting down all activities below the relevant scale. This is just what one would expect from a system which cannot sustain free low-momentum color excitations.

Thus what we find is perfectly in line with what one usually expects the physics of this theory to be.

  Suppose then that there is a jump in the low momentum limit of the gluon propagator $D(k, h)$ at $h = 0$.  This raises the question: Are there states that analytic or numerical calculations at $h = 0$ miss?\footnote{This question does not arise for the Gribov-type propagator, for which the order of limits commutes.}  This question is prompted by what happens in a ferromagnetic spin lattice.  Recall that to find the spontaneously magnetized state, one adds an arbitrarily small external magnetic field, $h$, which is then taken to zero.  Of course in the ferromagnetic case it is the first derivative $w'(h)$ that is discontinuous at $h = 0$ whereas in our case it is the second derivative $w''(h)$, so the two cases are different, and we do not have the  answer to this question.  Moreover, although the source $J_\mu^a(x)$ breaks Lorentz and color invariance, we do not expect these symmetries to be spontaneously broken in the gauge field theory case.  As a word of caution, we should note that the Landau gauge condition $k_\mu A_\mu = 0$ is not well defined at $k = 0$, and we are perhaps just seeing the result of a singular gauge choice.  However we have attempted to circumvent this possible problem by always taking the limit $k \to 0$ from finite $k$ in our analytic studies.  One possibility is that the different limits correspond to different gauge choices.  Another point that is (as yet) entirely not understood is what implication, if any this jump has for supplemental conditions to the Dyson-Schwinger equations, e.\ g.\ boundary conditions, which select among the solution manifold of the functional equations. Thus there are still some formal developments to be investigated.

Summarizing, we have understood quite a bit more about how Yang-Mills theory works. We have resolved the apparent discrepancies from different approaches to determining the gluon propagator.  We have found that they do not disagree, they just take limits in opposite order, and thus obtain consistently different results.

\appendix

\section{Locating the Gribov horizon} \label{locatinghorizon}

	The configuration $A$ lies on the (first) Gribov horizon when the lowest non-trivial eigenvalue $\lambda_0(A)$ of the Faddeev-Popov operator $M(A)$ vanishes.  For configuration (\ref{planebc}), the Faddeev-Popov eigenvalue problem reads
\beq
\{- \p^2  -[b + c \cos(k x_1)] \p_2 \ e_3 \times \} \psi_0(x) = \lambda_0 \psi_0(x),
\eeq
where $\psi_0(x)$ is a color vector, $e_3$ is an $x$-independent unit color-vector, and $\times$ is the bracket of the $su(2)$ Lie algebra.  To find the lowest eigenvalue, we take
\beq
\psi_0(x) = f(x) \ \eta,
\eeq
where $\eta = (e_1 -  i \sigma e_2)/ \sqrt 2$ is a color vector satisfying $e_3 \times \eta = i \sigma \eta$ with $\sigma = \pm 1$, and $f(x)$ is an ordinary function of position, so the eigenvalue equation reads
\beq
\{- \p^2  -  i \sigma [b + c \cos(k x_1)] \p_2  \} f(x) = \lambda_0 f(x).
\eeq
To proceed, we take
\beq
f(x) = \varphi(x_1) \exp(i p x_2), 
\eeq
where $p = 2 \pi m / L$, and $m$ is an integer, and the eigenvalue equation becomes one-dimensional,
\beq
[ - \p_1^2 + p^2 + \sigma pb + \sigma p c \cos(k x_1)] \varphi(x_1)  = \lambda_0 \varphi(x_1).
\eeq
The lowest eigenvalue is obtained when $\sigma = \pm 1$ is given by the sign function, $\sigma = - {\rm sign}(pb)$,
and we have
\beq
[ - \p_1^2 + p^2 - |pb| + q \cos(k x_1)] \varphi(x_1)  = \lambda_0 \varphi(x_1),
\eeq
where $q \equiv \sigma pc = - \rho |p| c$, and $\rho \equiv {\rm sign}(b)$.

This is a familiar one-dimensional Schr\"odinger eigenvalue problem with a periodic potential.  The lowest non-trivial eigenvalue is obtained if $|p|$ has the smallest non-zero value $|p| = 2 \pi /L$, and the eigenvalue problem reads
\beq
[ - \p_1^2 + q \cos(k x_1)] \varphi(x_1)  = \lambda' \varphi(x_1),
\eeq
where $\lambda' = \lambda_0 - p^2 + |pb|$, and $q = - (2\pi/L) \rho c$.  We are interested in the infinite-volume limit $L \to \infty$, with the momentum $k$ held fixed.  This corresponds to $q \to 0$, and we may treat the term in $q$ as a small perturbation.  In zeroth order we have $\varphi(x_1) = 1$, and the exact solution is of the form
\beq
\varphi(x_1) = 1 + \sum_{m = 1}^\infty a_m \cos(mkx_1),
\eeq
where $a_m = O(q^m)$.  
In the limit $L \to \infty$, it is sufficient to take 
\beq
\varphi(x_1) = 1 + a_1 \cos(kx_1),
\eeq
and systematically ignore higher terms.  With this expression for $\varphi$, the eigenvalue equation reads
\beq
k^2 \cos(kx_1) a_1 + q \{ \cos(kx_1) + (1/2)[ 1 + \cos(2kx_1)] a_1 \} = \lambda' [ 1 +a_1 \cos(k x_1) ].
\eeq
We drop the higher-order term $\cos(2kx_1)$, and obtain.
\beq
(1/2) q a_1 = \lambda'  \hspace{3cm} k^2 a_1 + q = \lambda' a_1,
\eeq  
which gives
\beq
\lambda' = (1/2) [ k^2 \pm (k^4 + 2 q^2)^{1/2} ],
\eeq
and $a_1 = 2 \lambda'/q$.  The lower sign gives the lower eigenvalue, and to lowest order in $q$ we have $\lambda' = - q^2 / 2 k^2 = - p^2 c^2/2 k^2 $, or
\beq
\lambda_0(b, c) =  - |pb| + p^2 \Big( 1 - { c^2 \over 2 k^2 } \Big),
\eeq
and $a_1 = - q/k^2 = \rho |p| c/k^2 $, with wave function
\beq
\label{ground state}
\psi_0(x) = [ 1 + (\rho |p| c / k^2) \cos(kx_1) ] \exp(ipx_2) (e_1 - i \sigma e_2)/\sqrt 2.
\eeq
where $\sigma = - {\rm sign}(pb)$ and $\rho = { \rm sign}(b)$.  The lowest non-trivial eigenvalue is obtained by setting $|p| = 2\pi/L$, which gives~(\ref{lambda0}) and~(\ref{psi0}).

 \section{Bounds from trial wave functions}

For any configuration inside the Gribov region, $A \in \Omega$, the Faddeev-Popov operator $M(A)$ is positive, $(\psi, M(A) \psi) \geq 0$, for any trial wave function $\psi$,
\beq
(\p_\mu \psi, \p_\mu \psi) - (\psi, A \times \p_\mu \psi) \geq 0.
\eeq
We write this as
\beq
(J, A) \leq (\p_\mu \psi, \p_\mu \psi),
\eeq
where
\beq
\label{trialJ}
J_\mu^b(x) \equiv [f^{abc} \psi^{a*}(x) \p_\mu \psi^c(x)]^{\rm tr},
\eeq
and the superscript ``tr" means that the transverse part is taken.  This bound gives the inequality
\beqa
\exp W(J) & = &   \int_\Omega dA \ \rho(A) \  \exp(J, A) 
\nonumber \\
& \leq &   \int_\Omega dA \ \rho(A) \  \exp(\p_\mu \psi, \p_\mu \psi) 
\nonumber \\
& = & \exp(\p_\mu \psi, \p_\mu \psi),
\eeqa
and we obtain the bound, for any trial wave function $\psi$,
\beq
W(J) \leq (\p_\mu \psi, \p_\mu \psi),
\eeq 
where $J$ is given in (\ref{trialJ}).  

Of course this bound is not optimal in general.  However if $A$ lies on the Gribov horizon, and if the trial wave function $\psi$ is the exact wave function belonging to the lowest non-trivial eigenvalue of the Faddeev-Popov operator $M(A)$, then the bound just given for any trial wave-function becomes the optimal bound.  To see this, observe that if $A \in \p \Omega$, then $(\psi, M(A) \psi) = 0$, which is the same as $(J, A) = (\p_\mu \psi, \p_\mu \psi)$, where $J$ is given in (\ref{trialJ}).  In this case, the bound $W(A) \leq (\p_\mu \psi, \p_\mu \psi)$ may be written $W(J) \leq (J, A)$.  Moreover $J$, given in (\ref{trialJ}), is normal to the Gribov horizon, as we have seen previously (\ref{lagrangeeq}).  Thus all the conditions for the optimal bound are satisfied.

As an example, take
\beq
\psi = \exp(i p \cdot x) \eta
\eeq
where $\eta = (e_1 - i e_2)/\sqrt 2$, and $e_1, e_2, e_3$ form an orthonormal basis in color space.  This yields the bound
\beq
p_\mu \int d^dx \ A_\mu^3 \leq p^2 V.
\eeq
By choosing $p_\mu = (2\pi/L) \delta_{\mu \nu}$ we get, for each $\nu$,
\beq
\int d^dx \ A_\nu^3 \leq (2\pi/L) V.
\eeq
This agrees with the optimal bound (\ref{constconfig}).


\end{document}